%% file: main.tex
\documentclass[acmsmall]{acmart}
\AtBeginDocument{%
  \providecommand\BibTeX{{%
    \normalfont B\kern-0.5em{\scshape i\kern-0.25em b}\kern-0.8em\TeX}}}

\setcopyright{acmlicensed}
\acmJournal{POMACS}
\acmYear{2022} \acmVolume{6} \acmNumber{3} \acmArticle{47} \acmMonth{12} \acmPrice{15.00}\acmDOI{10.1145/3570607}

\received{August 2022}
\received[revised]{October 2022}
\received[accepted]{November 2022}



\newcommand*\circled[1]{\raisebox{.5pt}{\textcircled{\raisebox{-.5pt}{\footnotesize #1}}}}

\newcommand*\varcircled[1]{\raisebox{.5pt}{\textcircled{\raisebox{-0.2pt}{\hspace{-.5pt}\scriptsize #1}}}}

\usepackage{hyperref}
\usepackage{tikz}
\usepackage{amsmath}
\allowdisplaybreaks[4]
\usepackage{verbatim}
\usepackage{makecell}
\usepackage{pifont}
\usepackage{subfigure}
\usepackage{listings}
\usepackage{booktabs}
\usepackage{diagbox}
\usepackage{enumitem}
\usepackage{balance}
\usepackage{subfigure}
\usepackage{xspace}
\newcommand{\cone}[0]{\ding{192}\xspace}
\newcommand{\ctwo}[0]{\ding{193}\xspace}
\newcommand{\cthree}[0]{\ding{194}\xspace}
\newcommand{\cfour}[0]{\ding{195}\xspace}
\newcommand{\cfive}[0]{\ding{196}\xspace}

\newcommand{\FP}{\textsc{FuncPipe}\xspace}
\newcommand{\LambdaML}{LambdaML\xspace}
\newcommand{\Hybrid}{HybridPS\xspace}
\newcommand{\LambdaGA}{LambdaML-GA\xspace}
\newcommand{\HybridGA}{HybridPS-GA\xspace}
\newcommand{\Tarnawski}{TPDMP\xspace}
\newcommand{\Bayes}{Bayes\xspace}

\newcommand{\bs}{batch size\xspace}
\newcommand{\para }[1]{\medskip \noindent  {\bf #1}}
\newcommand{\1}{{\em (i)}}
\newcommand{\2}{{\em (ii)}}
\newcommand{\3}{{\em (iii)}}

\begin{document}

\title[FuncPipe: A Pipelined Serverless Training Framework]{FuncPipe: A Pipelined Serverless Framework for Fast and Cost-Efficient Training of Deep Learning Models}

\author{Yunzhuo Liu}
\email{liu445126256@sjtu.edu.cn}
\affiliation{%
  \institution{Shanghai Jiao Tong University}
  \country{China}
}

\author{Bo Jiang}
\email{bjiang@sjtu.edu.cn}
\thanks{Bo Jiang is the corresponding author.}
\affiliation{%
  \institution{Shanghai Jiao Tong University}
  \country{China}
}

\author{Tian Guo}
\email{tian@wpi.edu}
\affiliation{%
  \institution{Worcester Polytechnic Institute}
  \country{U.S.}
}

\author{Zimeng Huang}
\email{lukehuang@sjtu.edu.cn}
\affiliation{%
  \institution{Shanghai Jiao Tong University}
  \country{China}
}

\author{Wenhao Ma}
\email{mwh1233@sjtu.edu.cn}
\affiliation{%
  \institution{Shanghai Jiao Tong University}
  \country{China}
}

\author{Xinbing Wang}
\email{xwang8@sjtu.edu.cn}
\affiliation{%
  \institution{Shanghai Jiao Tong University}
  \country{China}
}

\author{Chenghu Zhou}
\email{zhouch@lreis.ac.cn}
\affiliation{%
  \institution{Chinese Academy of Sciences}
  \country{China}
}


\begin{abstract}
\input{tex/abstract}

\end{abstract}

\maketitle

\input{tex/introduction}

\input{tex/background}
\input{tex/design-system}

\input{tex/evaluation}

\input{tex/related_work}

\input{tex/conclusion}

\input{tex/ack}

\newpage
\bibliographystyle{ACM-Reference-Format}
\bibliography{references}

\newpage
\appendix

\input{tex/notation}

\input{tex/model_app}

\input{tex/linearization}

\input{tex/api_app}

\input{tex/eval_model_accuracy}

\end{document}

%% file: tex/abstract.tex

Training deep learning (DL) models in the cloud has become a norm.
With the emergence of serverless computing and its benefits of true pay-as-you-go pricing and scalability, systems researchers have recently started to provide support for serverless-based training. 
However, the ability to train DL models on serverless platforms is hindered by the resource limitations of today's serverless infrastructure and DL models' explosive requirement for memory and bandwidth. This paper describes \FP, a novel pipelined training framework specifically designed for serverless platforms that enable fast and low-cost training of DL models.
\FP is designed with the key insight that model partitioning can be leveraged to bridge both memory and bandwidth gaps between the capacity of serverless functions and the requirement of DL training. 
Conceptually simple, we have to answer several design questions, including how to partition the model, configure each serverless function, and exploit each function's uplink/downlink bandwidth.
In particular, we tailor a micro-batch scheduling policy for the serverless environment, which serves as the basis for the subsequent optimization. Our Mixed-Integer Quadratic Programming formulation automatically and simultaneously configures serverless resources and partitions models to fit within the resource constraints. Lastly, we improve the bandwidth efficiency of storage-based synchronization with a novel pipelined scatter-reduce algorithm.
We implement \FP on two popular cloud serverless platforms and show that it achieves 
7\%-77\% cost savings and 1.3X-2.2X speedup
compared to state-of-the-art serverless-based frameworks.

%% file: tex/introduction.tex
\section{Introduction}


Serverless computing has recently been exploited for distributed training as an alternative to traditional VM-based training~\cite{siren, cirrus, lambdaml, lambdadnn}.
Serverless-based training has many attractive properties.
First, it relieves machine learning (ML) practitioners from management tasks such as configuring VMs' environment and setting up distributed training clusters~\cite{cirrus,siren}.
Second, its true pay-as-you-go pricing helps ML practitioners avoid paying for idle resources, e.g., during the trial-and-error process of model training~\cite{siren}. 
Such trial-and-error processes can last a long time: based on our analysis of two popular DL training traces, Philly~\cite{philly} and Helios~\cite{helios}, users spent more than half of the end-to-end training time on this process.
Third, it exhibits good resource elasticity and can auto-scale to many \emph{workers}, i.e., serverless functions~\cite{siren,dorylus}. The increased parallelism is especially beneficial for DL training, e.g., the ability to launch many workers for fast hyperparameter tuning and the flexibility to terminate workers for early-stopped configurations~\cite{hyper-tuning-ns-1, hyper-tuning-ns-2}.



However, today's cloud serverless platforms, e.g., AWS Lambda, impose stringent limits on available memory and bandwidth that make it difficult to utilize them to train resource-intensive DL models directly. 
Despite recent system efforts in \emph{enabling model training} on cloud serverless platforms~\cite{lambdadnn, lambdaml}, ML practitioners still do not have access to fast and cost-efficient serverless-based training. Our empirical analysis reveals the following two key challenges.

\begin{figure}[t]
    \centering

	\subfigure[\LambdaML performance.]{\label{fig:intro_comm}
	\begin{minipage}[t]{0.4\columnwidth}
			\centering
			\includegraphics[width=0.65\columnwidth]{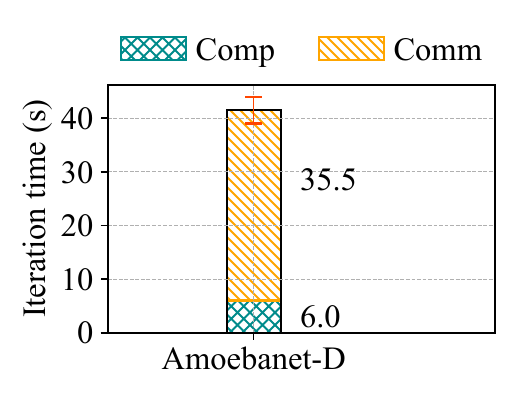}
		\end{minipage}
	}
	\subfigure[Training with three configurations.]{\label{fig:intro_policy}
	\begin{minipage}[t]{0.4\columnwidth}
			\centering
			\includegraphics[width=0.65\columnwidth]{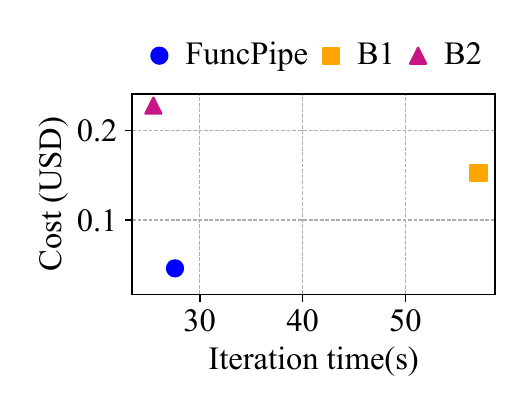}
		\end{minipage}
	}
    \caption{
    \textbf{(a)} LambdaML encounters communication bottleneck when training an AmoebaNet-D model.
    \textbf{(b)} Optimized model partition and serverless resource configurations greatly improve the overall performance.  
    }
    \label{fig:motivation}
\end{figure}

First, serverless functions have \textit{restricted communication capability} compared to traditional cloud VMs that does not meet the growing communication demand for training DL models. 
For instance, the maximum bandwidth of an AWS Lambda function is only about 70 MB/s (0.5 Gb/s)~\cite{serverless_bandwidth1, serverless_bandwidth2} while a VM can have up to 100 Gb/s bandwidth. Moreover, serverless functions lack the ability for \emph{direct inter-function communication}, which makes their communications rely on intermediaries such as Amazon S3 and ElastiCache~\cite{siren, lambdaml}.
Compounding with other training options like data parallelism, existing serverless-based training frameworks can suffer severe communication bottlenecks. Fig.~\ref{fig:intro_comm} shows the average iteration time for training a 900 MB \emph{AmoebaNet-D} with 8 AWS Lambda functions using \LambdaML~\cite{lambdaml}, a state-of-the-art serverless-based training framework. The computation takes only 6 seconds for each iteration, while communication takes nearly 6X of that. 

Second, serverless functions are allowed a much smaller memory footprint than traditional VMs, hindering their ability to achieve a cost-efficient \emph{computation-to-communication ratio}. 
For example, AWS Lambda offers up to 10 GB memory size for a serverless function~\cite{aws_lambda}, while a VM has up to TBs of memory. In contrast, the memory consumption during training can easily reach tens of GBs and increases with the model size and the \emph{activation size} which is proportional to the batch size.
We observe in Fig.~\ref{fig:intro_comm} that increasing the computation-to-communication ratio of the \emph{AmoebaNet-D} model from 0.17 to 0.45 (with local batch size 32) would require about 30GB of memory, far above the current memory cap of AWS Lambda functions.
Existing serverless-based training frameworks provide no effective solution to improve this low computation-to-communication ratio~\cite{lambdaml, cirrus, siren}.

Our work aims at improving the speed and cost-efficiency of training DL models on cloud serverless platforms. 
In designing \FP, we address the above challenges through two major approaches, \textit{utilizing model partition techniques} and \textit{improving storage-based communication efficiency}. 
Our key insight is that model partitioning is not only good for overcoming the memory constraint but also useful in relieving the communication burden in training. 
Through model partition, we can increase the computation to communication ratio by supporting a larger training batch size (e.g., 32 vs. 8 without partition for AmoebaNet-D model) on each serverless function. Model partition also reduces the size of gradients, compared to data parallelism, on each serverless function but at the cost of additional communication, i.e., exchanging outputs between different partitions. Because these outputs are much smaller than the gradients, the total amount of data transfer with model partition is still a small portion of data parallelism-based training.
On the other hand, to further speed up the function-storage communication, we design a new scatter-reduce algorithm for synchronization that pipelines the upload and download tasks. Our pipelined scatter-reduce design \emph{simultaneously} utilizes both uplink and downlink bandwidth of serverless functions, a desirable feature not supported by LambaML, recent work for serverless-based training~\cite{lambdaml}.

At the core, \FP explores pipeline parallelism~\cite{gpipe, dapple, pipedream, megatron}, a type of parallel structure based on the model partition, for \emph{fast and low-cost} serverless-based training.
We answer two key design questions:
\1 how to partition the DL model for the pipeline;
and \2 how to allocate resources for each serverless function.
Though the first question has been widely studied in server-based pipeline training~\cite{dapple, pipedream, nips_partition}, it poses a more complicated optimization question in a serverless environment. Specifically, those prior work often assume static training resources, i.e., a fixed number of workers with fixed resources, with the goal to only maximize training throughput.
In contrast, \FP simultaneously determines the model partition, the number of replicas for each partition (hence the number of workers) and the resource configuration for each worker, with a large search space (i.e., a large number of workers and many possible resource configurations), to optimize for both training throughput and cost.

Fig.~\ref{fig:intro_policy} compares the performance of training an AmoebaNet-D with the model partition and serverless resource allocation configurations found by \FP and two existing algorithms (denoted by \textsc{B1} and \textsc{B2}).
We can see that an efficient configuration can greatly improve the overall performance.
Training with configuration found by \FP decreases 52\%/70\% iteration time/cost compared to \textsc{B1}, and reduces cost by 80\% compared to \textsc{B2} with only 8\% time overhead. 
However, it is nontrivial to identify these effective configurations for different models because the decisions for model partition and resource allocation are \emph{tightly-coupled}. The optimal model partition depends on the allocated resources, and the training performance achieved by the model partition determines whether the resource allocation is cost-effective. Therefore, a joint decision of the two aspects is required, making the optimization problem more challenging.

In short, we make the following main contributions. 

\begin{itemize}
    \item We design and implement \FP; a novel pipelined serverless framework that enables fast and cost-efficient training of DL models with layered structures. \FP provides user-friendly Python APIs that require minimal changes to user code. We make the source code of \FP publicly available~\footnote{\url{https://github.com/liu445126256/FunPipe}}.
    
    \item We propose a novel pipelined scatter-reduce algorithm that utilizes uplink and downlink bandwidth during model synchronization. Our algorithm reduces the synchronization time by 6\%-26\% and the overall iteration time by 2\%-18\% compared to the non-pipelined scatter-reduce used in \LambdaML~\cite{lambdaml}.

    \item We formulate a co-optimization problem of model partition and resource allocation using Mixed-Integer Quadratic Programming (MIQP). 
    Our optimization approach finds configurations that achieve an average of 80\% faster training speed or 55\% lower cost compared to existing approaches~\cite{nips_partition, bayes}.
    
    \item We conduct an extensive evaluation of \FP on two popular serverless platforms with representative DL models.
    \FP achieves 1.3X-2.2X training speedup and 7\%-77\% cost reduction, compared to \LambdaML, the state-of-the-art serverless training framework~\cite{lambdaml}.
\end{itemize}

%% file: tex/background.tex
\section{Background}
\label{sec:background}

\subsection{Serverless Computing}
Serverless computing provides a new paradigm for deploying applications. 
To use serverless computing on major platforms such as AWS Lambda~\cite{aws_lambda}, users upload their applications (including code and dependencies) and execute them as stateless serverless functions. 
Though serverless users can execute the functions and obtain the computation results without managing the underlying computing infrastructures, users need to configure the functions with the proper amount of resources.
The task of resource configuration in today's serverless platforms amounts to deciding the memory allocation; given a memory allocation, other resources like CPU and network bandwidth are allocated accordingly by the cloud providers. 
Further, users are charged proportionally to the allocated memory and the actual runtime of their applications.

Serverless computing makes it easy to launch many instances of the same serverless function (up to thousands) concurrently; each function instance is often ready to run within seconds or even milliseconds~\cite{cold_start1, cold_start2}.
Serverless provides the true \emph{pay-as-you-go} pricing models and has garnered interests from both industry and academia~\cite{serverless_application1, serverless_application2, serverless_application3, serverless_application4} to run event-driven workloads such as in-memory caching~\cite{Wang2020-to,Romero2021-ly,llama} and workloads that benefit from a high degree of parallelism, including distributed training~\cite{siren,cirrus,lambdadnn, lambdaml}.
While prior work focuses on \emph{enabling} distributed DL training on the serverless platforms, this work improves the training speed and cost-efficiency with approaches including pipelining and co-optimization of model partition and resource allocation.

\subsection{Distributed Training}
\label{sec:bk_distributed}


Distributed training refers to training a machine learning model with multiple workers that communicate over different networks~\cite{distributed_network, distributed_nvlink}. 
When using distributed training, DL practitioners need to make two major decisions, i.e., determining how to divide tasks among workers (parallelism) and how to communicate progress (synchronization).

\para{Parallelism.} \emph{Data parallelism}~\cite{data_parallelism1, data_parallelism2} is a widely adopted type of parallelism where each worker maintains a replica of the entire model and a portion of the dataset. 
In a \emph{training iteration}, i.e., the processing of one batch of data, workers calculate gradients on their local data and then communicate the gradients to update the model parameters. 
Another type is \emph{model parallelism}~\cite{model_parallelism2, model_parallelism3, model_parallelism4} where the model is partitioned across workers. 
Rather than compute the gradients for the entire model, each worker will only compute the data batch on the assigned partition and then communicate the output to the worker that holds the next partition. 
Consequently, model parallelism often leads to reduced memory consumption and communication data size on each worker, as the total size of transferred data is usually much smaller than that of model gradients in data parallelism.
Model parallelism can be further combined with data parallelism by having multiple replicas for each model partition~\cite{hybrid_parallelism1, hybrid_parallelism2, hybrid_parallelism3}. 
In such a case, workers working on the same partition need to communicate gradients. Similar to model parallelism, such hybrid parallelism reduces communication as only gradients for partitions are exchanged when compared to data parallelism.
Our work falls under the general hybrid parallelism as we leverage \emph{pipeline parallelism}, detailed in \S\ref{bg:pipeline}, where we partition the model and allow data parallelism for each partition.

\para{Synchronization.} Distributed training can either use synchronous~\cite{gpipe, chimera, dapple, gems} or asynchronous protocols~\cite{pipedream,asynchronous1,asynchronous2} to instruct when workers can proceed to work on the next data batch. Synchronous protocol, in essence, ensures that workers work on the same version of model parameters by aggregating gradients from all workers to update the model at the end of every iteration. Therefore, it is not subject to potential accuracy convergence issues faced by asynchronous training. In this work, we focus on synchronous training to avoid impact on converged accuracy.

\para{Serverless-based distributed training.} In serverless-based training, each worker is mapped to a running serverless function. Existing serverless-based training frameworks~\cite{siren, cirrus, lambdaml, lambdadnn} are based on data parallelism and differ mainly in their communication designs.
There are two major communication architectures, i.e., centralized and decentralized. Parameter Server (PS) is a typical centralized architecture where workers upload their gradients to a central server, and from whom fetch the latest updated model parameters~\cite{li2019cmdare_icdcs,li2021syncswitch_icdcs}. A recently proposed serverless-based training framework Cirrus~\cite{cirrus} adopts such an architecture. In decentralized architecture, workers communicate with each other following the steps of specific communication algorithms, such as all-reduce~\cite{all_reduce1, all_reduce2, all_reduce3, ringallreduce} and scatter-reduce~\cite{all_reduce1}. The state-of-the-art serverless-based training framework \LambdaML~\cite{lambdaml} adopts decentralized architecture and proposes a storage-based scatter-reduce method to combat the performance degradation due to indirect communication via the intermediary storage.
Our work also uses decentralized architecture as it is shown to have better scalability in general~\cite{lambdaml}.
The major difference between our work and \LambdaML is that our work explores more complicated pipeline parallelism as the key to addressing the performance bottleneck of serverless-based training. We focus on combining pipeline parallelism with serverless design and optimizing the performance of serverless-based pipeline training.

\subsection{Pipeline Training and Model Partition}
\label{bg:pipeline}

Pipeline parallelism has been explored to improve the resource utilization of traditional server-based model parallel distributed training ~\cite{megatron, gpipe, dapple, pipedream, chimera}.
At a high level, pipeline parallelism divides a data batch into micro-batches and treats each model partition as a stage in the pipeline. 
During the training, micro-batches will be scheduled to go through the model partitions in a pipelined fashion to simultaneously utilize resources of different stages.
As such, pipeline parallelism can address model parallelism's low resource utilization problem by reducing worker idle time.

One of the key designs to ensure the efficiency of pipeline training is model partition.
Although model partition is a well-studied topic in model parallelism~\cite{socc_partition_model, icml_partition_model, midware_partition_model, nips_partition_model, iclr_partition_model}, the proposed algorithms usually achieve sub-optimal performance when applied to pipeline training due to mismatched goals. The goal for model parallelism is to minimize the processing time of one batch, while in pipeline training, the goal is to minimize the processing time of multiple micro-batches in an iteration. 
Prior work on server-based pipeline training has proposed several model partition strategies to improve pipeline training throughput~\cite{dapple, pipedream, nips_partition}.
However, they often assume static training resources, i.e., a fixed number of workers with a fixed amount of resources.
In serverless computing, we are presented with the flexibility to scale up to many workers and to easily configure workers with different amounts of resources.
Such flexibility is a double-edged sword: it gives us more knobs to improve the performance and reduce the monetary cost, while it also makes the problem of configuring pipeline parallelism more difficult.
In this work, we tackle the challenge of effectively partitioning the model in pipeline training and configuring resources in a serverless environment to achieve high training throughput and incur low cloud bills.

%% file: tex/design-system.tex
\begin{figure*}[t]
\center
\includegraphics[width=.95\linewidth]{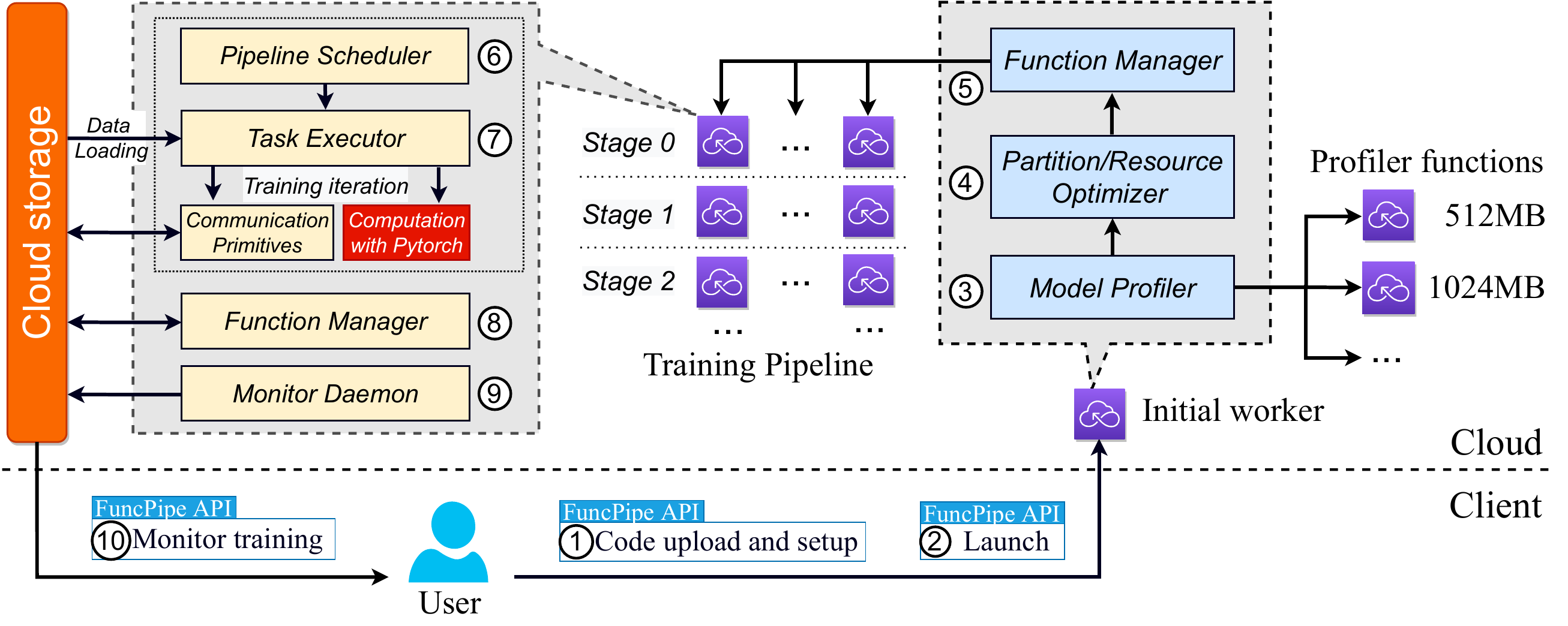}

\vspace*{-.5em}
\caption{
\textbf{\FP system architecture and workflow.} 
The two gray boxes enclose \FP components.
The blue blocks are the startup components active in the initial worker, and the yellow blocks are the runtime components in a training worker.}
\label{fig:funcpipe_workflow}
\end{figure*}

\section{\FP Design}

In this section, we present \FP, a novel pipelined serverless framework for efficiently training DL models. 
\S\ref{sec:sys_overview} provides an overview of the system architecture and workflow.
\S\ref{sec:design_pipeline} and \S\ref{sec:design_scatter_reduce} give detailed designs of the training pipeline and pipelined scatter-reduce, respectively. 
\S\ref{sec:design_coopt} presents our co-optimization approach for model partition and resource allocation.

\subsection{Overview} 
\label{sec:sys_overview}

\para{System architecture.}
As shown in Fig.~\ref{fig:funcpipe_workflow},
\FP consists of three parts, startup components, runtime components, and client-side APIs. The startup and runtime components are displayed in the two gray boxes in the figure, represented by blue and yellow boxes, respectively. Those components run on the serverless platform and interact with cloud storage and client-side APIs. The client-side APIs enable the users to set up, launch, and monitor the training with minimum effort. Our choice of cloud storage as function-to-function communication channel follows the recent work  \LambdaML~\cite{lambdaml}. Specifically, we choose object storage, e.g., AWS S3, for its low monetary cost. Even though in-memory storage like Elasticache and DynamoDB provides lower access latency, they are often more costly. Plus, latency has little impact on the performance of serverless-based training, whose communication bottleneck is often caused by the limited function bandwidth.



\para{Workflow.}
The workflow of \FP is shown in Fig.~\ref{fig:funcpipe_workflow}.
The user first prepares the training code using \FP APIs and then sets up and launches training from the client side (\circled{1} and \circled{2}). 
In the beginning, an initial worker with the startup components performs the preparation work: \circled{3} \emph{Model Profiler} profiles \emph{model layers}, i.e. network topologies in the architecture of the deep learning model such as convolutional layers and fully connected layers, on serverless functions with different memory allocations; \circled{4} With the gathered layer-wise information, e.g., computation time, parameter and activation size, \emph{Partition/Resource Optimizer} finds the optimal model partition and the best resource allocation based on our MIQP formulation  (\S\ref{sec:design_coopt}); \circled{5} \emph{Function Manager} configures resources and launches all training workers to start the pipeline. 

When the pipeline training starts, micro-batches are scheduled to traverse the pipeline with the help of the following components. \circled{6} Each worker runs a \emph{Pipeline Scheduler} and the scheduler decides the processing order of the micro-batches. 
\circled{7} \emph{Task Executor} handles the processing tasks by interacting with underlying storage-based \emph{Communication Primitives} and \emph{Pytorch}. It properly overlaps communication and computation. 
\circled{8} Each worker runs a \emph{Function Manager}, and the managers exchange information during training to ensure the health of the pipeline. As serverless functions have a limited lifetime, e.g., 15 minutes in AWS Lambda, \emph{Function Manager} checkpoints and restarts the worker at a designated time interval to avoid function timeout. The same procedure is adopted by prior work~\cite{cirrus, lambdaml}. 
Finally, \circled{9} \emph{Monitor Daemon} gathers and uploads training information that users can access using client-side API (\varcircled{10}).   




\begin{figure*}[t]
    \centering
    \includegraphics[width=\linewidth]{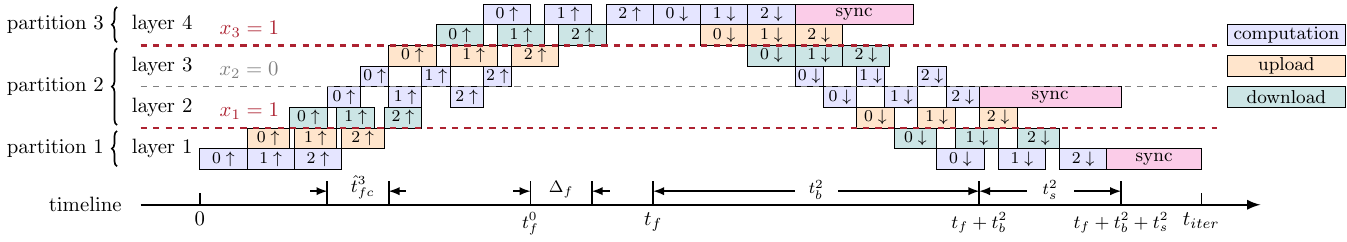}
    \caption{\textbf{Training pipeline of \FP.} Each block represents a processing task. The labeled index indicates the micro-batch that each block corresponds to, and the labeled up/down arrows represent the forward/backward processes respectively. Blocks in the vertical direction can overlap each other in execution.
    The notations on the timeline are used in the formulation in \S\ref{sec:design_coopt}.
    }
    \label{fig:pipeline}
\end{figure*}

\subsection{Training Pipeline}
\label{sec:design_pipeline}


We illustrate the pipeline design of \FP through the example in Fig.~\ref{fig:pipeline}. 
\FP uses the pipeline
to perform synchronous training that avoids potential convergence and accuracy issues.
\FP partitions the model and places each partition on a serverless worker. 
In a training iteration, the data batch is divided into micro-batches,
and they are scheduled to traverse the partitions in the following order:
\1 all micro-batches go through each partition to perform forward computation;
\2 after all forward computations have finished, the micro-batches go in a reversed order for backward computation, i.e., backpropagation. 

Each worker in our pipeline generally handles two types of tasks,  computation and communication. Communication tasks are further divided into \emph{upload}, \emph{download}, and \emph{sync}. The output of the partitions is communicated through \emph{upload} and \emph{download} to/from the cloud storage; \emph{sync} is required at the end of a training iteration if multiple workers are configured for a partition (i.e., data parallelism). It can be performed once the backward computation of the partition is completed.

Our micro-batch scheduling policy is similar to the one used by  GPipe~\cite{gpipe}, which was designed for server and GPU-based training. Our scheduling policy has two differences: it treats communication tasks as a pipeline stage and overlaps it with the computation task, and it uses a pipelined scatter-reduce algorithm (\S\ref{sec:design_scatter_reduce}) to utilize both uplink and downlink bandwidth for the \emph{sync} task.
Our communication-oriented optimization is driven by the key difference between serverless and server-based pipelines, i.e., the proportion of communication time in the overall training time. For example, in the server-based case, communication time is usually negligible as its workers can have large bandwidth, e.g., 100Gb RDMA or 300GB NVlink. In the serverless case, however, communication can take up a large proportion as serverless functions have limited bandwidth. 
More concretely, both \emph{upload} and \emph{download} times can be comparable to \emph{computation} time. The \emph{sync} time can even be significantly longer depending on the degree of data parallelism.

Other micro-batch scheduling policies~\cite{gems, dapple, chimera} could also be used but will lead to more complex pipeline structures and therefore introduce additional complexity in developing the co-optimization approach (\S\ref{sec:design_coopt}). In other words, we choose the current scheduling policy for its simplicity, and we leave exploring other scheduling methods as future work. 

\begin{figure*}[t]
\center
\includegraphics[width=\linewidth]{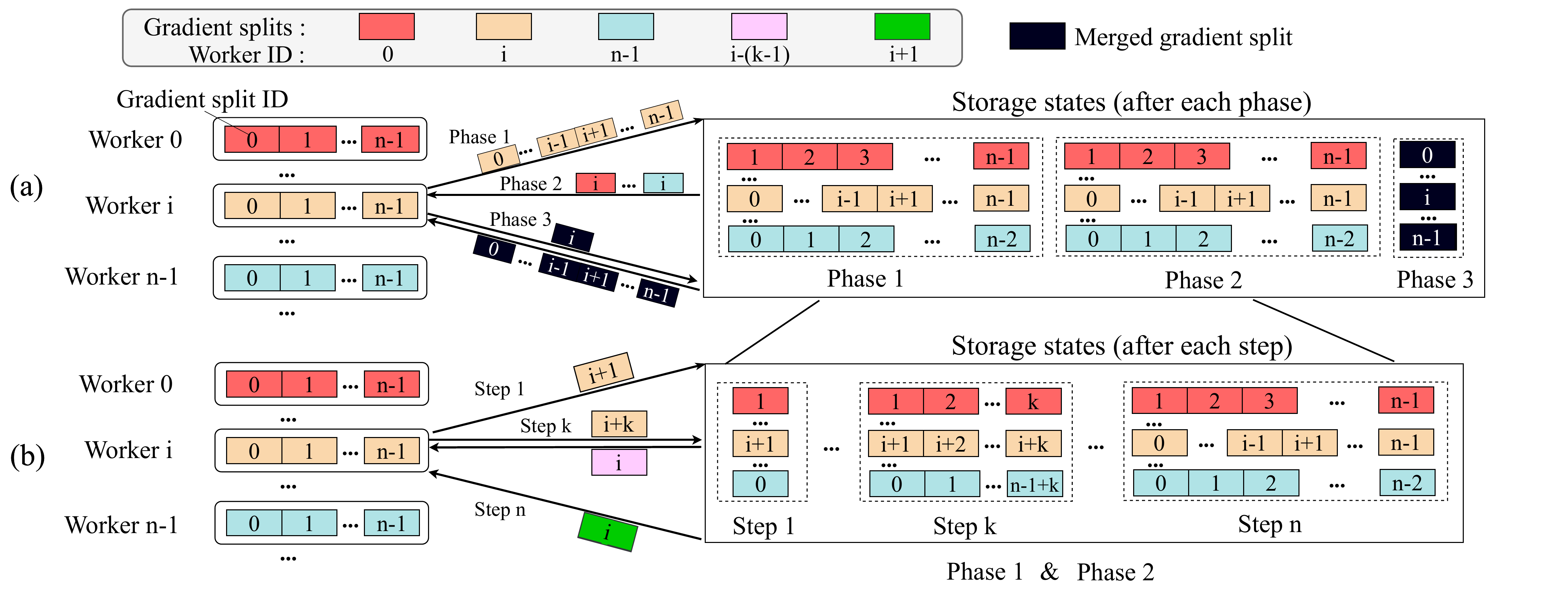}
\caption{\textbf{Illustration of our pipelined scatter-reduce.}
\textbf{(a)} The scatter-reduce in \LambdaML~\cite{lambdaml} has three phases where download and upload are performed serially. \textbf{(b)} Our pipelined scatter-reduce performs download and upload in duplex in \emph{phase 1} and \emph{phase 2}.
}
\label{fig:pipelined_scatter_reduce}
\end{figure*}

\subsection{Pipelined Scatter-Reduce}
\label{sec:design_scatter_reduce}
We identify one of the causes for the low communication efficiency in existing serverless-based training frameworks~\cite{lambdaml} as that the current storage-based synchronization design
fails to make efficient use of the network bandwidth. 
To address the problem, we propose a pipelined storage-based scatter-reduce method that simultaneously utilizes downlink and uplink bandwidth.
Fig.~\ref{fig:pipelined_scatter_reduce}(a) displays the state-of-the-art storage-based scatter-reduce method proposed in LambdaML~\cite{lambdaml}. 
It utilizes the computation resource of all workers for gradient aggregation by dividing the gradients as $n$ splits, where $n$ is the number of workers, and each worker is in charge of merging one split. The scatter-reduce process can be divided into three phases: in \textit{phase 1}, each worker uploads the $n-1$ gradient splits that other workers are in charge of to the storage. In \textit{phase 2}, the $i$-th worker retrieves all the $i$-th splits uploaded by other workers and computes the merged gradients. In \textit{phase 3}, each worker uploads its merged split and retrieves all other merged splits. 
The communication time of \textit{phase 1} and \textit{phase 2} are both $\frac{s_{grad}(n-1)}{n\cdot w}+t_{lat}$, where $s_{grad}$ is the size of the gradients, $w$ is the bandwidth of a worker and $t_{lat}$ is the latency for accessing the storage. The communication time of \textit{phase 3} is $\frac{s_{grad}}{w}+2t_{lat}$, and the total synchronization time is
\begin{equation}
    \label{eq:old_scatter}
    \begin{aligned}
        3\cdot \frac{s_{grad}}{w} - \frac{2s_{grad}}{n\cdot w} + 4t_{lat}.
    \end{aligned}
\end{equation}
As the upload in \textit{phase 1} and the download in \textit{phase 2} are performed serially, the network resource is not efficiently utilized. 

Our scatter-reduce further pipelines \textit{phase 1} and \textit{phase 2} to improve communication efficiency. The pipelined phase includes a total of $n$ steps, as shown in Fig.~\ref{fig:pipelined_scatter_reduce}(b):
\begin{itemize}
	\item In step $1$: worker $i$ uploads gradient split $i + 1$
	to storage. 
	\item In step $k$, for $2\leq k \leq n-1$: worker $i$ uploads gradient split $i+ k$ to storage while downloading split $i$ uploaded by worker $i -(k-1)$.
	\item In step $n$: worker $i$ downloads gradient split $i$ uploaded by worker $i+1$. 
\end{itemize}
We use arithmetic modulo $n$ in the above.
The communication time of each of the above steps is $\frac{s_{grad}}{n\cdot w} + t_{lat}$, and the time for $n$ steps is $\frac{s_{grad}}{w} + n\cdot t_{lat}$. The total synchronization time is
\begin{equation}
    \label{eq:pipelined_scatter}
    \begin{aligned}
        2\cdot \frac{s_{grad}}{w}+(2+n)t_{lat}.
    \end{aligned}
\end{equation}
Comparison of \eqref{eq:old_scatter} and \eqref{eq:pipelined_scatter} shows that the pipelined scatter-reduce achieves a noticeable reduction in the transfer time, i.e. from $3 \frac{s_{grad}}{w}  - \frac{2s_{grad}}{n\cdot w}$ to $2\frac{s_{grad}}{w}$. For example, for an AWS Lambda function with 70MB/s bandwidth, the data transfer time of synchronizing a 280MB model among 8 workers can be reduced by 27\%, from 11s to 8s.
Although our design can suffer higher latency with the increase of workers, the latency is much smaller than the data transfer time, e.g., the measured  $t_{lat}$ is less than $40$ms for AWS Lambda.

\input{tex/model}

\section{Implementation}


\FP is implemented on top of \emph{Pytorch} with 4012 lines of Python code. 
It provides easy-to-use APIs and requires minimal changes to legacy training code on the user side. A code example for using \FP is given in Appendix~\ref{app:api_example}.
\FP currently supports two serverless platforms, AWS Lambda and Alibaba Cloud Function Compute, and can be easily extended to other platforms 
as the platform API design in \FP is decoupled from the underlying SDK implementations, e.g., \emph{boto3} for AWS Lambda and \emph{fc2} for Alibaba Cloud Function Compute.

\para{Pipeline task overlap.} 
The different tasks, \emph{upload}, \emph{download}, and \emph{computation} have internal dependencies and different resource requirements, i.e., downlink bandwidth, uplink bandwidth, and CPU. 
These tasks are organized as Directed Acyclic Graphs (DAGs) and  handled by different threads in the \emph{Task Executor}.
Tasks of different types are processed in parallel; each is assigned a unique ID and contains a set of IDs representing its dependencies. 
A task is immediately processed once its dependencies are satisfied.

\para{Communication collectives.} 
\FP performs storage-based communications, including \emph{send-and-receive} between different partitions and \emph{scatter-reduce} among partition replicas.
The data communicated are serialized with the python library \textit{pickle} and 
uploaded to the storage bucket as files. Metadata information is included in the file name to distinguish different pairs and types of communication. 
Workers periodically query the cloud storage bucket to check for \emph{download}. 

\para{MIQP solution.} For models with over a hundred layers, solving the MIQP problem can take hours or even days, limiting its practical usage. 
As many model layers can have small memory consumption and short computation time, they can be merged with other layers to reduce the value of $L$, i.e. the total number of layers in optimization. 
By merging the layers, our method ensures a minute-level solution time.
Currently, we provide three options for the merging criterion, computation time, parameter size, or activation size. 
For all the tested models, merging by balancing the computation time achieves better performance and is adopted in our experiments. 

\FP provides two implementations for \emph{Partition/Resource Optimizer}.
The first one solves the MIQP optimization using serverless functions. 
However, off-the-shelf solvers can have licence limits that require additional support in order to be used in the serverless environment. 
For example, Gurobi requires user to have a Gurobi token server that grants temporary license~\cite{floating_license}. 
For users that want to avoid such effort, we provide a second implementation that solves the MIQP optimization at the client side. 
The information obtained by \emph{Model Profiler} is retrieved by the client for optimization and the results are uploaded back to the initial worker.    

\para{Limitation discussion.} 
Currently \FP does not support training  models that contains a layer that exceeds the maximum memory for a serverless function. A solution to this problem, and also a possible direction for further optimization is to use tensor parallelism~\cite{tensor_parallelism1, tensor_parallelism2, hybrid_parallelism1, megatron, megatron2}. Tensor parallelism partitions a tensor along specific dimensions, for example, Megatron~\cite{megatron} partitions transformer layer by splitting its weight matrix and FlexFlow~\cite{hybrid_parallelism1} partitions CNN layer in terms of both channels and input length. The major benefit we expect from using tensor parallelism is more fined-grained partition decision, which can potentially lead to more cost-efficient resource choices and the avoid of memory overflow caused by super large-sized layers. However, using tensor parallelism increases the complexity of the proposed co-optimization approach, as the extra decision dimensions greatly expand the search space. In such case, techniques like approximation may be required to make the solution of the MIQP practical. We leave extending \FP to tensor parallelism as future work.




%% file: tex/model.tex

\subsection{Co-optimization of Model Partition and Resource Allocation }
\label{sec:design_coopt}

To make the training pipeline fast and cost-efficient, we need to optimize the partition plan that splits model layers into different pipeline stages and the resource allocation for each stage. This plan includes the number of workers used for intra-stage data parallelism as well as the memory size of each worker. A major challenge here is the strong coupling between model partitioning and resource allocation, which defies most existing solutions that optimize only one aspect \cite{pipedream, dapple, nips_partition}. In this section, we formulate the co-optimization of model partition and resource allocation as a mixed-integer quadratic program. 

\subsubsection{Formulation of Optimization Problem}\label{sec:opt}

Consider a model with $L$ layers. 
 Let $\mathcal{D}=\{D_1, \dots, D_K\}$ be the set of possible degrees of data parallelism, where $D_{1}=1$, meaning no data parallelism.  Let $\mathcal{M}=\{M_1, \dots, M_J\}$ be the set of different memory sizes for serverless workers.
We use a binary variable $x_i$ to indicate whether the model is partitioned after layer $i$. Let $d\in \mathcal{D}$ be the degree of data parallelism. We enforce the same degree of data parallelism for all stages to reduce the problem complexity.  Let $m_i\in \mathcal{M}$ be the memory size of workers holding layer $i$. We parameterize $d$ and $m_i$ as $d = \sum_{k=1}^K y_k D_k$ and $m_i = \sum_{j=1}^J z_{i,j} M_j$ with binary variables $y_k$ and $z_{i,j}$, where $y_k = 1$ if $d = D_k$ and $z_{i,j} = 1$ if $m_i = M_j$. 
The number of micro-batches per worker is given by
 $\mu = \frac{M}{d}= \sum_{k=1}^K y_k \frac{M}{D_k}$, where $M$ is the total number of micro-batches.
Other notations will be introduced as needed; see Appendix \ref{app:notation} for a full table of notations. 

Our goal is to choose $(x_i)$, $(y_k)$ and $(z_{i,j})$ to minimize the  cost $c_{iter}$ and  time $t_{iter}$ per training iteration. We formulate it as the \emph{nonlinear binary integer program} in \eqref{prob:opt}, which we explain below. 
\begin{subequations}\label{prob:opt}
\begin{align}
\min \quad & \alpha_1\cdot c_{iter} + \alpha_2\cdot t_{iter} \label{eq:obj}\\
s.t. \quad & \mu \hat a_{i}+  \hat s_i(4-2y_{1}) + s_0 \leq m_i, \quad & 1 \leq i \leq L; \label{eq:mem}\\
&|m_i - m_{i-1}| \leq x_{i -1}\cdot M_{\max}, \quad & 2 \leq i \leq L; \label{eq:consistency}\\
&  \sum_{k=1}^{K} y_k=1, \quad \sum_{j=1}^{J} z_{i,j}=1, \quad &1 \leq i \leq L; \label{eq:unique}\\
& x_i, y_k, z_{i,j} \in \{0,1\}, & \forall i, j, k. \label{eq:binary}
\end{align}
\end{subequations}

The expressions for $c_{iter}$ and $t_{iter}$ will be given in \S\ref{sec:performance_model}.
We combine the two objectives into a single objective in \eqref{eq:obj} using the weighted sum method. Each pair of weights $(\alpha_1, \alpha_2)$ yields a Pareto optimal solution. 
As the weights vary, the solutions will trace out the Pareto Frontier \cite{pareto}.

To explain the constraints, we first introduce the hat operator. Given any sequence $u_1, u_2, \dots, u_L$ where $u_i$ is a quantity associated with layer $i$, we define
\begin{equation}\label{eq:forward_cs}
\hat u_{1} = u_{1}, \quad\hat u_i = u_i+ \hat u_{i-1} (1-x_{i-1}), \quad 2 \le i \le L,
\end{equation}
where $x_i$ are our decision variables for model partition. 
The hat operator accumulates quantities forwardly in each partition.
Let $\mathcal{H}$ denote the set of the highest layers of the partitions. For the example in Fig. \ref{fig:pipeline}, $\mathcal{H}=\{1,3,4\}$. For $i\in \mathcal{H}$,  $\hat u_i$ is the sum of the quantity $u_j$ over the partition containing layer $i$.  In Fig. \ref{fig:pipeline}, $\hat u_3 = u_2 + u_3$ is the sum over partition 2. 
  
The constraints \eqref{eq:mem} specify that the  memory consumption of each partition does not exceed the allocated memory of the corresponding worker. Let $s_i$ denote the parameter size and $a_i$ the activation size per micro-batch at layer $i$. 
For $i \in \mathcal{H}$,  $\mu\hat a_i$ is the memory for activations of $\mu$ micro-batches in the partition that layer $i$ belongs to; $\hat s_i(4-2y_{1})$ comprises three parts of memory consumption, $\hat s_{i}$ for parameters,  $\hat s_{i}$ for gradients, and $2(1-y_1)\hat s_{i}$ for serialized data during model synchronization. Note synchronization is needed only if $y_1 = 0$. The quantity $s_0$ is the basic memory consumption of a serverless worker, e.g., memory consumed by the framework. We only need the constraints for $i\in \mathcal{H}$, as the others are redundant. The constraints \eqref{eq:consistency} enforce consistency of the memory allocation for adjacent layers if they belong to the same partition, as they actually share the same workers, i.e. $m_i=m_{i-1}$ if $x_{i-1}=0$. With $M_{\max} = \max_{1\le j\le J} M_J$ being the maximum memory available,  the constraint for $i$ becomes vacuous when the model is partitioned after $i-1$, i.e. $x_{i-1}=1$. The constraints \eqref{eq:unique} and \eqref{eq:binary} specify that we choose exactly one degree of data parallelism and exactly one memory configuration for each layer.

To solve \eqref{prob:opt}, we convert it into an MIQP using standard linearization techniques, which is then solved by off-the-shelf solvers, e.g., Gurobi\cite{gurobi}. The details for linearization is in Appendix~\ref{sec:linearization}.

\subsubsection{Performance Model}\label{sec:performance_model}

\phantom{.}

\para{Iteration cost.} Recall the memory allocated for layer $i$ workers is $m_i = \sum_{j=1}^J z_{i,j} M_j$. Since layers of the same partition are assigned to the same workers, we only count the layers in $\mathcal{H}$, and the total memory of all workers is 
\begin{equation}
\label{eq:mem_cost}
c_{mem} = d\sum_{i\in\mathcal{H}} m_i = d \left(\sum_{i=1}^{L-1} x_{i}m_{i} + m_L\right)
\end{equation}
The cost of serverless functions is proportional to the product of their running time and memory allocation, so the iteration cost $c_{iter}$ is
\begin{equation}
\label{eq:money_cost}
\begin{aligned}
c_{iter} = P\cdot t_{iter}\cdot c_{mem}
\end{aligned}
\end{equation}
where $P$ is the unit price specified by the service provider.

\para{Iteration time.} 
As shown in Fig. \ref{fig:pipeline}, the iteration time $t_{iter}$ is given by
\begin{equation}\label{eq:iter_time}
t_{iter} 
= t_f + \max_{1\le i\le L} (t_b^i + t_s^i),
\end{equation}
where $t_f$ is the forward time. When layer $i$ is the lowest layer of a partition (e.g., layer 2 in Fig. \ref{fig:pipeline}),  $t_b^i$ is the backward computation completion time of that partition, and $t_s^i$ the corresponding model synchronization time. For other layers (e.g., layer 3 in Fig. \ref{fig:pipeline}),  
their sum $t_b^i+t_s^i$ will be dominated by that of the lowest layer of the same partition (e.g., layer 2), and hence their inclusion in \eqref{eq:iter_time} does not affect $t_{iter}$. 
Next we introduce the formulas for $t_f$, $t_b^i$ and $t_s^i$ in detail.

\paragraph{Forward and backward time.} We only show the calculation of the forward time $t_f$. The calculation of the backward time $t_b^i$ is similar and relegated to Appendix \ref{app:model}.  The forward time $t_f$  is
\[
t_f = t_f^0 + (\mu-1) \Delta_f,
\]
where $t_f^0$ is the time for the first micro-batch to traverse the  forward pipeline, $\Delta_f$ the lag between consecutive micro-batches at the end of the  forward pipeline,  and $\mu$ the number of micro-batches per worker. 
The time $t_f^0$ is given by
\[
t_f^0 = \sum_{i=1}^{L} t_{fc}^{i} +\sum_{i=1}^{L-1}  (t_{fu}^{i} + t_{fd}^{i}),
\]
where $t_{fc}^i$ is the forward computation time of layer $i$, $t_{fu}^i$ the upload time of the output of layer $i$ to the storage, and $t_{fd}^i$  the download time of the output of layer $i$ from the storage to layer $i+1$. 
 The individual terms are related to $(z_{i,j})$ by
\begin{align} 
t_{fc}^{i} &= \beta  \sum_{j=1}^{J} z_{i,j}T_{fc}^{i,j}, & 1\le i\le L,\nonumber \\ 
\label{eq:forward_layer2} t_{fu}^{i} &= x_i\left(\sum_{j=1}^J z_{i,j} \frac{o_i}{W_j}+ t_{lat}\right), & 1\le i\le L-1, \\
t_{fd}^{i} &= x_i\left(\sum_{j=1}^J z_{(i+1),j}\frac{o_{i}}{W_j}+ t_{lat}\right), & 1\le i\le L-1, \nonumber
\end{align}
where $T_{fc}^{i,j}$ is the forward computation time of layer $i$ by a worker with memory $M_j$, $\beta\ge 1$ is the average slowdown factor due to resource contention when we overlap computation and communication, $o_i$ is the output size of layer $i$, $W_j$ is the bandwidth of a worker with memory $M_j$, and  $t_{lat}$ is the measured latency to storage. The values of $T_{fc}^{i,j}$, $\beta$, $W_j$ and $t_{lat}$ are measured by the \emph{Model Profiler} during initial profiling. Note that communication times $t_{fu}^i$ and $t_{fd}^i$ are nonzero only if $x_i = 1$, i.e. there is a partition boundary after layer $i$.

The lag $\Delta_f$ is the maximum time of all stages, i.e.
\begin{equation*}
\Delta_f = \max \left\{\hat t_{fc}^{1:L},   t_{fu}^{1:(L-1)},   t_{fd}^{1:(L-1)} \right\},
\end{equation*}
where $t^{i_1:i_2}$ denotes the set of variables $t^i$ for $i_1\le i \le i_2$, and $\hat t_{fc}^i$ is related to $t_{fc}^i$ by \eqref{eq:forward_cs}. For $i\in \mathcal{H}$, $\hat t_{fc}^i$ is the computation time for the stage containing layer $i$.
For the example in Fig. \ref{fig:pipeline}, $\hat t_{fc}^3$ is the time for the second computation stage, 
consisting of layer 2 and layer 3. Note we only need to include $\hat t_{fc}^i$ for $i\in \mathcal{H}$, but the inclusion of the other $i$ gets rid of $\mathcal{H}$.

\paragraph{Synchronization time.}

When $i$ is the lowest layer of a partition, e.g., layer $2$ for partition 2 in Fig. \ref{fig:pipeline}, the synchronization time of that partition is
\begin{equation}\label{eq:sync_time}
t_{s}^{i} = (1 - y_{1})\left(\sum_{j=1}^J z_{ij} \frac{\tilde s_{i}}{W_j}\cdot \gamma + t_{lat} \cdot \delta\right),
\end{equation}
where $\gamma$ and $\delta$ are parameters that depend on the synchronization algorithm. For the \emph{pipelined scatter-reduce}, we have $\gamma = 2$ and $\delta = 2+d$ by \eqref{eq:pipelined_scatter}. The tilde operator is similar to the hat operator in \eqref{eq:forward_cs}, except that it accumulates the quantities backwardly so that $\tilde s_{i}$ of the lowest layer equals the size of the partition.
The model update time is negligible and hence not included. Note $t_{s}^{i}$ is positive only if the degree of data parallelism is more than $1$, i.e. $y_1 \neq 1$. When $i$ is not the lowest layer of a partition, we also define $t_s^i$ by \eqref{eq:sync_time}. The inclusion of those quantities do not affect the value in \eqref{eq:iter_time}, since $t_s^i \ge t_s^{i'}$ if $i'\ge i$ and layers $i$ and $i'$ belong to the same partition.

%% file: tex/evaluation.tex
\section{Evaluation}
This section first presents the overall performance of \FP by comparing it with state-of-the-art serverless-based training designs (\S\ref{sec:overall-perf}) and discusses its system scalability (\S\ref{sec:scalability}). We then validate the effectiveness of \FP's designs with component-wise study, including the evaluation of our pipelined scatter-reduce algorithm (\S\ref{sec:scatter_perf}) and co-optimization of model partition and resource allocation (\S\ref{sec:mip_perf}). Next, we discuss the effect of resource availability on different serverless platforms (\S\ref{sec:resource_availability}). Finally, we evaluate the performance of \FP with increased network bandwidth  (\S\ref{sec:future_proof}).

\subsection{Methodology}

\para{Cloud serverless testbed.} Our evaluation uses two of the mainstream serverless platforms, AWS Lambda~\cite{aws_lambda} and Alibaba Cloud Function Compute~\cite{ali_function_compute}, that provide different resource options. AWS Lambda provides a maximum of 10 GBs of memory allocation for each serverless function. Its corresponding cloud storage service, \emph{S3}, grants unlimited bandwidth to concurrent access. Alibaba Cloud Function Compute has different resource availability compared with AWS Lambda. It allows a maximum memory allocation of 32 GBs, and its cloud storage \emph{OSS} puts a limit on the concurrent bandwidth, e.g., a total of 10 Gb/s for a normal customer. Most of our evaluations are on AWS Lambda, and we leverage Alibaba Cloud Function Compute to study the impact of resource availability on different serverless platforms. 

\para{Models and datasets.} The DL models used for our evaluation are in Table \ref{tab:models}. \emph{ResNet101}, \emph{AmoebaNet-D18}, and \emph{AmoebaNet-D36} are  popular Convolution Neural Network (CNN)  models for computer vision tasks. 
\emph{BERT-Large} is a transformer model for natural language processing. 
We use the popular image classification dataset \emph{CIFAR-10} to train the CNN models.
To train \emph{BERT-Large}, we run masked language modeling on the dataset \emph{Wikitext-2}. 
We use synchronous Stochastic Gradient Descent (SGD) optimizer with 
the same global batch size (further explained in \S\ref{sec:overall-perf}) for all tested designs in the evaluation, and we report the average per-iteration training time and cost.

\begin{table}[t]
    \sf
    \centering
    \scriptsize
    \begin{tabular}{lrr}
    \toprule
	 \textbf{Model name} & \makecell[c]{\textbf{Parameter size} \textbf{(MB)}} &  \makecell[c]{\textbf{Activation size}  \textbf{per sample (MB)}} \\
    \midrule
      ResNet101  & 170   & 198  \\
      AmoebaNet-D18 & 476 &  432 \\
	 AmoebaNet-D36 & 900 &  697 \\
	 BERT-Large& 1153 &  263\\
    \bottomrule
    \end{tabular}
    \caption{
        \textbf{Models used for evaluation.} 
           \emph{AmoebaNet-D18} and \emph{AmoebaNet-D36} are two \emph{AmoebaNet-D} models with 18 and 36 normal cell layers, respectively. Both have a filter size of 256.
    }
    \label{tab:models}
 \end{table}

\para{Baselines.}
We compare \FP with existing serverless-based training designs with two different structures: the pure serverless-based structure and the hybrid PS structure (as introduced in \S\ref{sec:bk_distributed}). 
LamdaML~\cite{lambdaml} is the state-of-the-art pure serverless-based training framework, and it also includes an implementation of the hybrid design exemplified by Cirrus~\cite{cirrus}, an end-to-end serverless framework for ML training. These two baselines are referred to as \textit{\LambdaML} and \textit{\Hybrid}. We further integrate \emph{gradient accumulation}, a commonly adopted technique for reducing the memory consumption in training~\cite{ga1, ga2, ga3}. The resulting baselines are referred to as \textit{\LambdaGA} and \textit{\HybridGA}, both serving as baselines that have reduced worker memory allocation and better-balanced computation to communication time ratio. 
The baselines and their resource allocation strategies are summarized as follows.
\begin{itemize}
    \item \textbf{\LambdaML} follows a pure serverless-based training design. It uses the maximum memory allocation and maximum local batch size within the memory limit for each worker. This strategy reduces the number of workers used for training with a given global batch size.
    \item \textbf{\Hybrid} follows a hybrid PS training design and requires the use of parameter servers. We select the instance with the lowest cost that can perform our tasks without incurring CPU or memory bottleneck at the parameter server, i.e., a \emph{c5.9xlarge} instance on AWS and a \emph{r7.2xlarge} instance on Alibaba. The resource allocation of workers follows the same strategy in \LambdaML~\cite{lambdaml} for a fair comparison. 
    Note that we replace the data serialization API in the implementation of \cite{lambdaml} with the python \texttt{pickle} module to utilize the worker network bandwidth better. 
    We observe that our modification improves training speed and cost. For example, before this modification, \Hybrid could only achieve a throughput of about 20 MB/s; the current implementation can fully utilize the bandwidth at about 70 MB/s.   
    \item \textbf{\LambdaGA} applies gradient accumulation to the \LambdaML baseline. It uses the same number of workers as \LambdaML but allocates \emph{the minimum memory} required after performing gradient accumulation for each worker. We use a batch size of 1 for each accumulation step to minimize memory consumption.
    \item \textbf{\HybridGA} uses a similar resource allocation strategy and the same batch size for each accumulation step as \LambdaGA.  
\end{itemize}

To validate the effectiveness of our co-optimization on model partition and resource allocation, we compare it with two existing algorithms. 
\begin{itemize}
    \item \textbf{\Tarnawski} is the latest graph-based model partition algorithm for server-based pipeline training~\cite{nips_partition}. It maximizes the pipeline training throughput with a fixed amount of resources. To apply \Tarnawski to the serverless scenario,
    we perform a grid search on the resource allocation and optimize the model partition with \Tarnawski for each allocation. We select the configuration that minimizes the objective function in \eqref{prob:opt}.
    \item \textbf{\Bayes} is a black-box optimization method that has been proved effective in deciding cloud configurations~\cite{cherrypick}. It generates a configuration, measures its performance, and iteratively refines the decision. \Bayes can be used to optimize the model partition and resource allocation jointly. However, with the large search space of our  problem, \Bayes can require many rounds of optimization just to find a feasible configuration. For example, it fails to find configurations that do not cause out-of-memory (OOM) errors for over half of our training tasks within 20 rounds of optimization. 
    To reduce the prohibitive time cost of real-world measurement, we evaluate each configuration with our performance model, which has a high accuracy of 88\% as shown in Appendix~\ref{app:model_accuracy}.
    Using performance models in place of the actual measurement is recently proposed and demonstrated to produce good optimization performance ~\cite{linnanlamcts,Zhao2021MultiobjectiveOB}. We run a total of 100 rounds of optimization to minimize the objective function in \eqref{prob:opt}.
    
\end{itemize}

\para{\FP settings.}
For the evaluation, we use $8$ discrete memory allocation choices, i.e., \emph{[512MB, 1024MB, 2048MB, 3072MB, 4096MB, 6144MB, 8192MB, 10240MB]}.
We empirically set the micro-batch size to $4$ as it achieves a generally better performance on the evaluation models.
We use four pairs of weights of $(\alpha_1, \alpha_2)$, i.e., \emph{$[(1,0),(1,2^{16}),(1,2^{19}),(1,2^{22})]$}, to locate the corresponding points on the Pareto Frontier. These weights are chosen empirically because they can generate results that represent very distinct speed and cost trade-offs. The same weights are used for the baseline algorithms \Tarnawski and \Bayes.    

\para{Recommendation.} \FP also recommends a configuration out of the optimized results, labeled as \emph{Recommendation} in subsequent figures. 
Denote by $t_{mc}$ and $c_{mc}$ the training time and cost of the minimum cost configuration obtained using weights $(1,0)$.
Assume the training time and cost of a given configuration as $t_{p}$ and $c_{p}$.
We use $\delta = (\frac{t_{mc}}{t_{p}} - 1)/(\frac{c_{p}}{c_{mc}} - 1)$ to represent how efficient a configuration is by comparing its speedup with its cost increase over the cheapest configuration. In our evaluation \FP recommends the fastest configuration that satisfies $\delta \geq 0.8$. 


	
	


\subsection{Overall Performance}
\label{sec:overall-perf}

\begin{figure*}[tb]
    \centering
	\subfigure[Batch size = 16]{\label{b16}
    \includegraphics[width= 0.95\columnwidth]{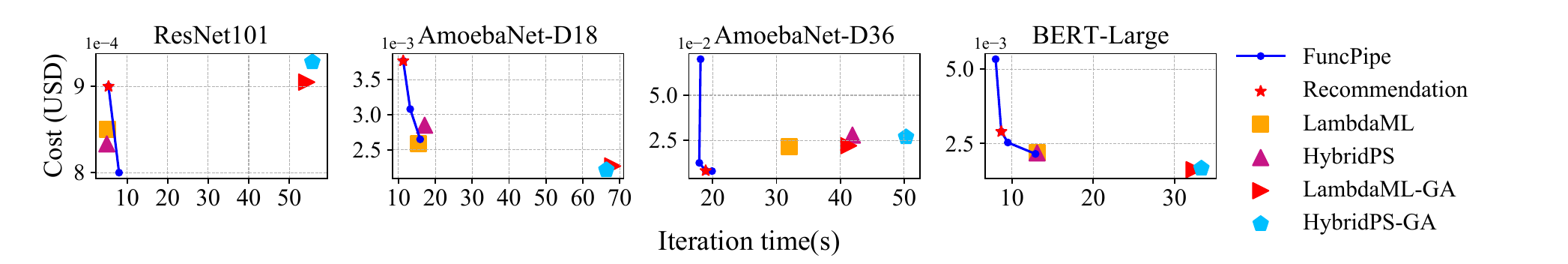}
	}
	
	\vspace*{-.5em}
	\subfigure[Batch size = 64]{\label{b64}
    \includegraphics[width=0.95\columnwidth]{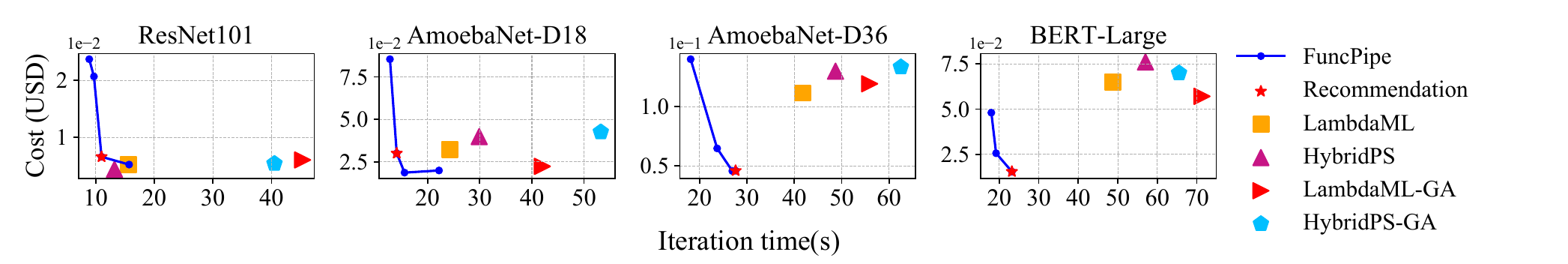}
	}
	
	\vspace*{-.5em}
	\subfigure[Batch size = 256]{\label{b256}
    \includegraphics[width=0.95\columnwidth]{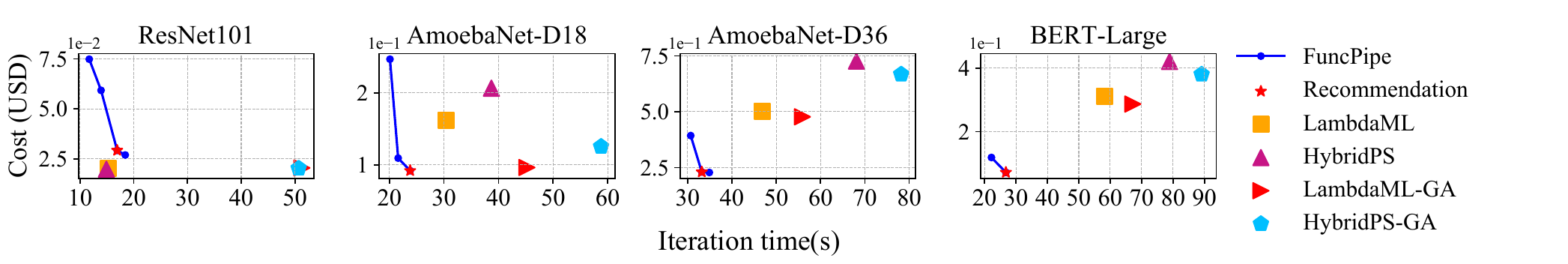}
	}

	\vspace*{-.5em}
    \caption{\textbf{Overall performance.} \FP outperforms existing designs in both training speed and cost in most of the test cases and achieves comparable or faster performance in other cases.
    }
    \label{fig:overall_perf}
\end{figure*}

The training performance of \FP and its comparison with existing serverless-based training designs are shown in Fig.~\ref{fig:overall_perf}. 
Generally, \FP achieves better performance in both training speed and cost over existing designs in most of the test cases (comparable or faster performance in other cases). And the performance improvement increases with the model size and global \bs.
The results are obtained with three commonly adopted global \bs of $16$, $64$, and $256$. The performance of each baseline method is represented as a single point in the figure, and for \FP, it is a curve consisting of the points corresponding to the configurations obtained using the four pairs of weights. 
Note that there can be fewer than four points on a curve as different weights may lead to the same configuration. The configuration recommended by \FP is also highlighted in this figure. 
We make the following key observations.

First, \FP achieves \textbf{1.3X-2.2X} training speedup and \textbf{7\%-77\%} cost reduction compared with the best-performing baseline \LambdaML when training \emph{AmoebaNet-D18}, \emph{AmoebaNet-D36} and \emph{BERT-Large} with global batch sizes of $64$ and $256$. The \textbf{2.2X} speedup and \textbf{77\%} cost reduction is achieved when training \emph{BERT-Large} with global batch size $256$. The improved training speed and cost-efficiency come from the reduced communication time and increased computation to communication ratio, as further illustrated in \S\ref{sec:time_breakdown}. 

Second, when training on a single worker is feasible, i.e., training with batch size 16, existing designs can achieve cost-efficiency similar to that of \FP since no communication overhead exists. However, our follow-up experiments show that their training speed cannot be further improved given more resources. This is because more resources change the training from a single worker to multiple workers, which incurs prohibitive communication costs for the existing designs. In contrast, \FP can achieve up to 1.6X speedup (training \emph{BERT-Large}) over the best-performing serverless-based training baseline (\LambdaML) when given more resources (2.4X cost).


Third, the hybrid design, \Hybrid, achieves comparable or even better performance than \LambdaML when training \emph{ResNet101}. 
However, with the increase in model size and global batch size (leading to the use of more workers), the server node in this centralized structure can be heavily burdened.
As a result, we can observe noticeable performance gap between \Hybrid and \LambdaML when training \emph{AmoebaNet-D36} and \emph{BERT-Large} in Fig.~\ref{b256}. 
In addition, we see that the use of gradient accumulation (\LambdaGA and \HybridGA) can reduce the training cost at the price of a longer training time. 
However, the reduction is neither significant nor guaranteed to exist. 
We attribute it to the use of gradient accumulation, which reduces the memory allocation but may incur higher costs due to the increased runtime.

\subsection{Training Time Breakdown}
\label{sec:time_breakdown}

\begin{figure}[t]
    \centering
	\subfigure[BERT-Large  (BS=16)]{\label{breakdown1}
	\begin{minipage}{0.24\columnwidth}
	\centering
    \includegraphics[width= 0.93\columnwidth]{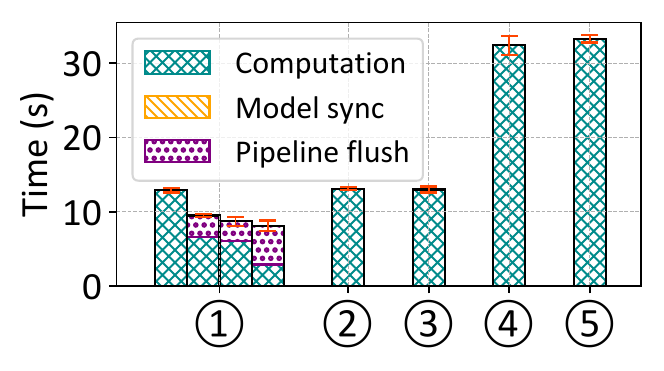}
    \end{minipage}
	}
	\subfigure[ResNet101  (BS=64)]{\label{breakdown2}
	\begin{minipage}{0.23\columnwidth}
	\centering
    \includegraphics[width=0.93\columnwidth]{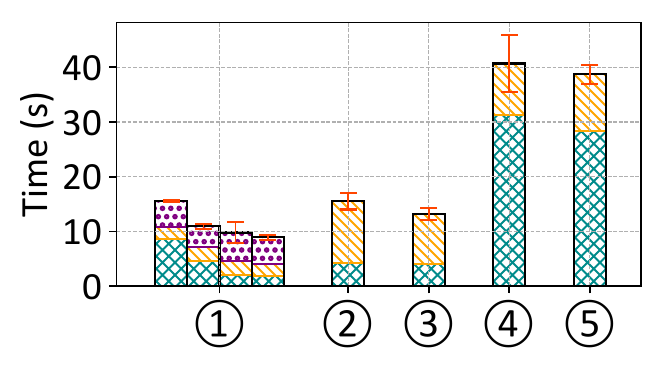}
    \end{minipage}
	}
	\subfigure[BERT-Large (BS=64)]{\label{breakdown3}
    \begin{minipage}{0.23\columnwidth}
	\centering
    \includegraphics[width= 0.93\columnwidth]{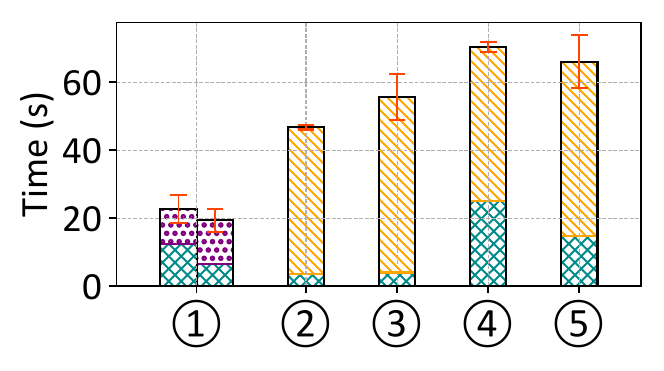}
     \end{minipage}
	}
	\subfigure[AmoebaNet-D36 (BS=64)]{\label{breakdown4}
    \begin{minipage}{0.23\columnwidth}
	\centering
    \includegraphics[width= 0.93\columnwidth]{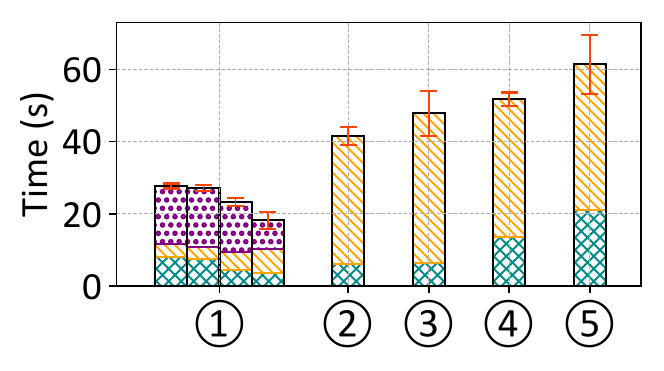}
     \end{minipage}
	}
	
	\vspace*{-.3em}
    \caption{\textbf{Training time breakdown.} 
    Labels: \cone {\small \FP}, \ctwo  {\small \LambdaML}, \cthree {\small \Hybrid}, \cfour {\small \LambdaGA},  \cfive {\small \HybridGA}.
    The multiple bars of {\small \FP} correspond to different configurations on the Pareto Frontier.
    Legend shared across figures.
    }
    \label{fig:time_breakdown}
\end{figure}

Fig.~\ref{breakdown1} displays the time breakdown for training \emph{BERT-Large} (Fig.~\ref{b16}). 
The small \bs allows the baseline methods to train the model on a single worker (no communication time) and thus fully utilize the computation resource and achieve cost-efficient training. 
However, their training speed cannot be improved any further. 
As their workers already have the maximum memory allocation, increasing the resource usage means using more workers. 
Such scaling up incurs high communication costs and stalls the training---synchronizing \emph{BERT-Large} (1153MB) with 70MB/s bandwidth can take tens of seconds, which is longer than the total computation time.
This shows that \FP can be faster than existing designs even when training with a small batch size.

Fig.~\ref{breakdown2} shows the time breakdown of training \emph{ResNet101} with \bs 64 (i.e., Fig.~\ref{b64}). 
The improvement in training speed achieved by \FP is relatively smaller than in Fig.~\ref{breakdown3} and Fig.~\ref{breakdown4}. 
This is because when the model size is small, the synchronization time of \LambdaML and \Hybrid can be close to the sum of pipeline flush time and intra-stage model synchronization time in \FP. 
This suggests that we expect small improvement or comparable performance from \FP with small models.

Figs.~\ref{breakdown3} and \ref{breakdown4} show the time breakdown for training \emph{BERT-Large} and \emph{AmoebaNet-D36}, respectively (i.e., Fig.~\ref{b64}). 
The breakdown shows that the performance improvement of \FP  in Fig.~\ref{b64} can be largely attributed to the reduced communication time, i.e., its pipeline flush time and intra-stage model synchronization time are much lower than the synchronization time of \LambdaML. 
We can also see that \FP has a larger computation to communication time ratio compared with the baseline methods, making \FP more cost-efficient.

\subsection{System Scalability}
\label{sec:scalability}

Next, we evaluate the scalability of \FP by comparing its performance to the best-performing design, i.e., \LambdaML, based on observations from \S\ref{sec:overall-perf}. 
For this experiment, we use the total amount of allocated memory to denote the system resource. 
Further, we use the global batch size to specify the amount of work. 
As such, we are evaluating both \FP and \LambdaML's ability to handle more work (i.e., increased global batch size) given more resources (i.e., total memory).
For \LambdaML, we increase the global batch size and resource usage by adding more workers.
Each worker is allocated the maximum memory and uses the maximum local batch size according to the resource strategy of \LambdaML. 
For \FP, we increase the global batch size and use the recommended configuration.

\begin{figure}[t]
\begin{minipage}[t]{0.48\linewidth}
    \includegraphics[width=1.02\linewidth]{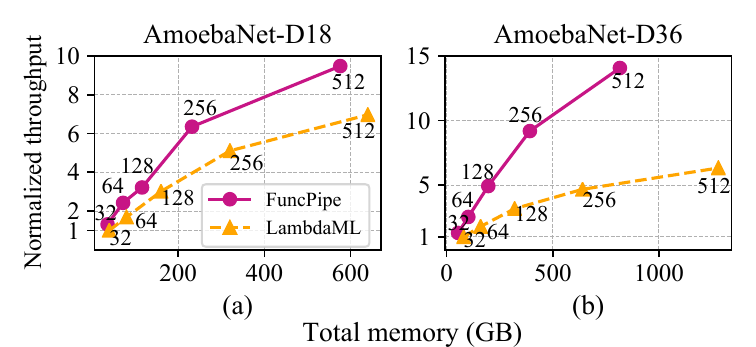} 
    \caption{\textbf{System scalability test.} 
    {\small\FP} achieves higher throughput and is more robust to bandwidth contention. 
    Each data point is annotated with the global batch size.} 
    \label{fig:scalability}
\end{minipage}%
    \hfill%
\begin{minipage}[t]{0.47\linewidth}
    \includegraphics[width=1.04\linewidth]{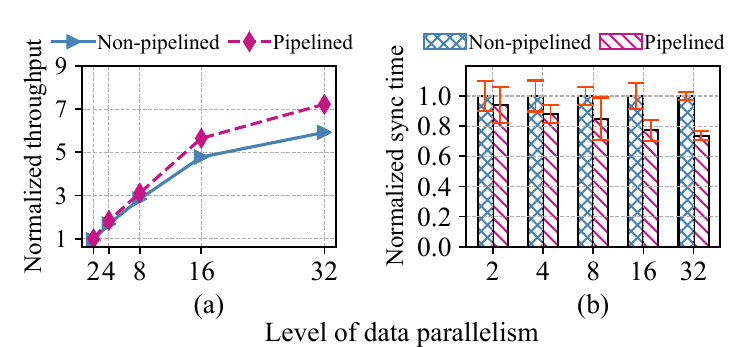} 
    \caption{\textbf{Performance of our pipelined scatter-reduce method.}
     (a)  Our design achieves 2\%-22\% higher training throughput and (b) 6\%-26\% lower synchronization time.} 
    \label{fig:sync_time}
\end{minipage} 
\end{figure}

Fig.~\ref{fig:scalability} reports the average training throughput, i.e., number of processed samples per second, on model \emph{AmoebaNet-D18} and \emph{AmoebaNet-D36}. 
The training throughput is normalized to that of \LambdaML with global \bs 32.
We first observe that \FP achieves higher training throughput than \LambdaML when given the same resource allocation. 
For example, when training the \emph{AmoebaNet-D36} model, the throughput is 180\% higher when both use 800 GB total memory.
Second, both \FP and \LambdaML exhibit a sublinear scaling up performance with \FP scaling better than \LambdaML.
We find that reduced per-worker network bandwidth causes the sublinear scaling up performance. 
The per-worker bandwidth reduction was also observed in prior work~\cite{dorylus}, and we suspect that it is because the serverless platforms schedule different serverless functions to the same machine, and thus they share a bandwidth capacity.
Additionally, we see that \FP is less affected by the bandwidth reduction than \LambdaML, possibly due to the effectiveness of \FP's designs in reducing the overall communication burden.

\subsection{Scatter-Reduce Communication Efficiency}
\label{sec:scatter_perf}
We compare our pipelined scatter-reduce design with \LambdaML's non-pipelined scatter-reduce~\cite{lambdaml}.
To perform the comparison, we use the recommended configuration for training \emph{AmoebaNet-D18} with a global batch size of $32$. 
The configuration divides the model into three stages, each with a data parallelism of 2. 
We gradually increase the level of data parallelism (the global batch size is increased proportionally) from $2$ to $32$ and compare the training throughput. 
As shown in Fig.~\ref{fig:sync_time}(a), the two scatter-reduce methods achieve similar performance with small data parallel levels at the beginning (pipelined scatter-reduce has 2\% higher throughput).
As the data parallel level increases, we observe a growing performance gap, and pipelined scatter-reduce achieves a 22\% higher training throughput than non-pipelined scatter-reduce. 
This increased performance gap can be understood in two ways. 
First, the increased data parallelism level increases the difference in transfer time of the two algorithms, as seen by comparing \eqref{eq:old_scatter} and \eqref{eq:pipelined_scatter}.
Theoretically, a reduction of up to 33\% in transfer time can be achieved.
Fig.~\ref{fig:sync_time}(b) shows that the gap between the synchronization time gradually increases from 6\% and reaches 26\%. 
Second, the increased data parallelism level uses more workers. 
Based on our observation in AWS Lambda, more workers can reduce the available bandwidth per worker.
As such, the communication time can take up a larger proportion of the overall training time, thus emphasizing 
the benefit of communication optimization.
In summary, our pipelined scatter-reduce can effectively improve communication efficiency. 


\begin{figure*}[!t]
    \center
    \includegraphics[width=0.95\linewidth]{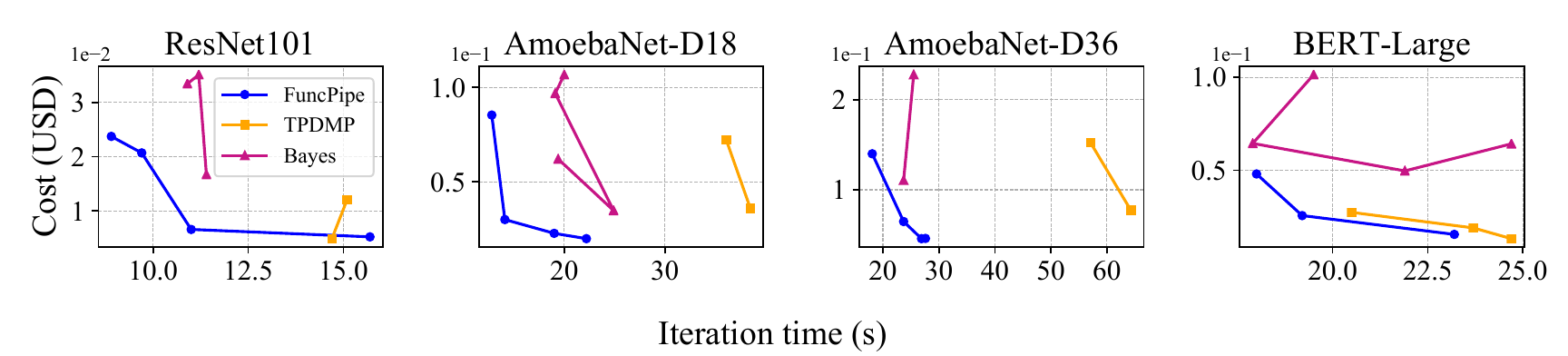}
    \caption{\textbf{Co-optimization performance evaluation.} 
    The global batch size is $64$. The performances with other batch sizes are similar.
    }
    \label{fig:mip_perf}
\end{figure*}

\subsection{Co-optimization Performance}
\label{sec:mip_perf}
We evaluate the performance of our co-optimization design by comparing it with existing model partition/resource allocation algorithms in terms of training performance and solution time.

\para{Training performance.} Fig.~\ref{fig:mip_perf} compares the model partition and resource allocation policies found by our co-optimization method and those by two existing algorithms~\cite{nips_partition, bayes}. Note that some methods in the figure contain fewer points as they generate the same configuration for different pairs of weights.
The results show that our design achieves the best overall performance. 
Compared to \emph{\Tarnawski}, our design has a comparable average training cost (within 3\% difference) but an average speedup of \textbf{1.8X} when optimized for the same objective function. 
The performance gap between our design and \emph{\Tarnawski} suggests the benefit of co-optimizing the model partition and resource allocation. 
Compared to \emph{\Bayes}, our co-optimization method achieves \textbf{7\%} higher average training speed and \textbf{55\%} lower average cost. 
We observe that the policies generated by \emph{\Bayes} often have higher monetary costs; 
we attribute \emph{\Bayes}'s cost-inefficiency to its tendency to over-provision the resource to avoid infeasible solutions, i.e., policies that lead to OOM error. 

\para{Solution time.} We evaluate the algorithms on the client side using an \textit{Intel(R) Core(TM) i5-10210U CPU}. The average solution time for each configuration in Fig.~\ref{fig:mip_perf} is \textbf{274s}, \textbf{603s}, \textbf{45s} for \FP, \Tarnawski and \Bayes respectively. The results show that \FP achieves the best performance with a reasonable solution cost, i.e., minute level. When the optimization problem is solved on the client side, it incurs no cloud bills; when it is solved in the cloud, such minute-level solution cost is negligible to the training cost.

\begin{figure}[!t]
    \center
    \includegraphics[width=1\linewidth]{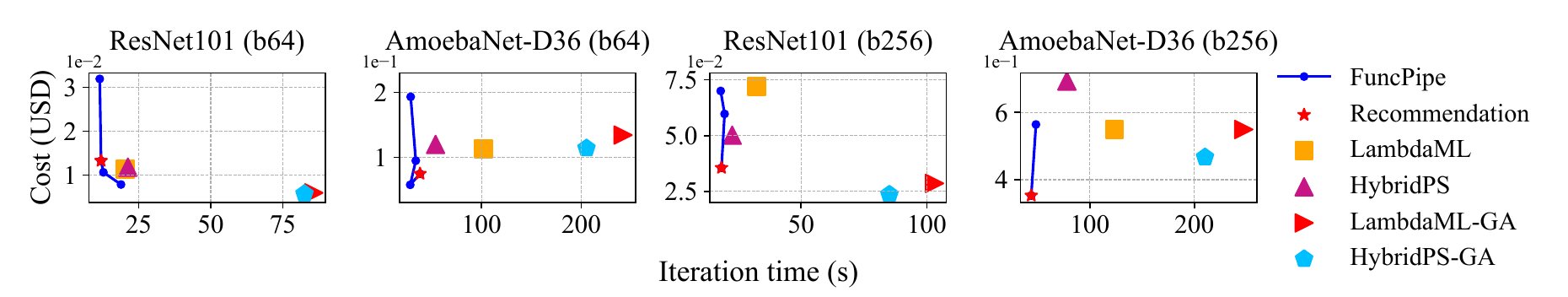}
    \caption{\textbf{Performance on Alibaba Cloud.} With the same limit on total communication bandwidth, \FP
    achieves up to \textbf{1.8X} speedup and \textbf{49\%} cost reduction compared with the best-performing baseline \Hybrid.
    }
    \label{fig:ali_perf}
\end{figure}

\subsection{Impact of Different Resource Availability}
\label{sec:resource_availability}

We evaluate the performance of \FP on the Alibaba Cloud to understand the potential impact of different resource availability. 
The major difference between AWS and Alibaba cloud is that the bandwidth of Alibaba Cloud storage OSS~\cite{ali_oss} has a total limit of 10Gb/s. The same bandwidth limit exists for the VM server used by the \Hybrid baseline. We study how the same bandwidth bottleneck affects the performance of these methods. Due to the space limit, we only show the results of training \emph{ResNet101} and \emph{AmoebaNet-D36} with global \bs 64 and 256 in Fig.~\ref{fig:ali_perf}. 
Overall, we find that \FP demonstrates similar benefits in Alibaba Cloud to AWS: comparable performance or small improvement on small-sized models and better performance in both training speed and cost as the model size and global \bs increase, with up to \textbf{1.8X} speedup and \textbf{49\%} cost reduction compared with the best-performing baseline \Hybrid. Note that the best baseline differs from AWS Lambda, as Alibaba cloud functions achieve higher throughput communicating with VM than with the object storage as we observe.
%
This result shows that \FP can alleviate the effect of the limited bandwidth.

Other platforms~\cite{azure_cloud} may have similar limits on the storage-side bandwidth, e.g., Azure Storage has a total limit of 25Gb/s~\cite{azure_storage}. Such storage-side bandwidth bottleneck may limit the ability of \FP to scale out, and \FP may eventually be outperformed by \Hybrid as the bandwidth of the latter can be increased by scaling up the parameter server. One solution for this is to use a VM-based storage design, like Pocket~\cite{pocket}, and the total bandwidth can be increased the same way as \Hybrid. In this case, we expect \FP to achieve better performance than \Hybrid, as the evaluation has demonstrated the performance benefits of \FP under the same bandwidth. Extending \FP to VM-based storage and further comparing to the \Hybrid design is left as future work.

\subsection{Impact of Increased Function Bandwidth}
\label{sec:future_proof}

\begin{figure}[!t]
    \center
    \includegraphics[width=1.0\linewidth]{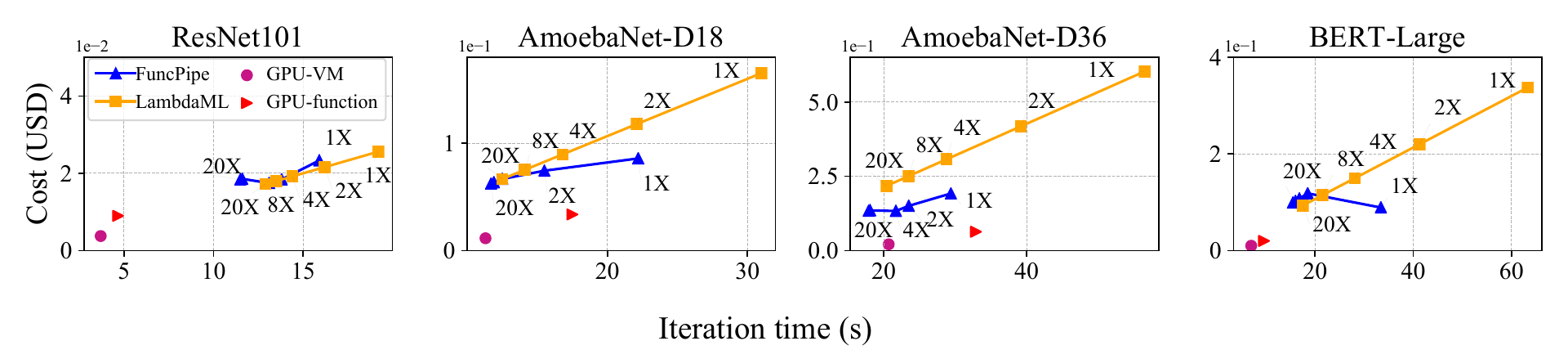}
    \caption{\textbf{Iteration time and cost with the increase of network bandwidth.} We gradually increase the bandwidth to 20X of the current function bandwidth. Note that each curve contains 5 points (some points are overlapped), and they correspond to the results with 1X, 2X, 4X, 8X, and 20X bandwidth, respectively. We also include a point of VM GPU-based training to demonstrate the performance gap with serverless CPU-based training. Such gap can be greatly narrowed once GPU is enabled for serverless function.  
    } 
    \label{fig:future_proof}
\end{figure}

As our breakdown analysis in \S\ref{sec:time_breakdown} shows that the improvement achieved by \FP mostly comes from the reduced communication time, we are interested in the performance of \FP when the network bandwidth increases. We simulate the performance of \FP with the performance model proposed in \S\ref{sec:design_coopt} by changing the value of bandwidth $W$. We compare the performance with that of the best-performing baseline \LambdaML, which is simulated using its analytical model~\cite{lambdaml}. 
Fig.~\ref{fig:future_proof} reports the training speed and cost as we gradually increase the bandwidth to 20x of the current function bandwidth in AWS Lambda, i.e., from about 0.5 Gb/s to 10 Gb/s, which is a common bandwidth for a VM. 

Generally, as the bandwidth increases, the performances of \FP and \LambdaML improve. 
The performance improvement of \LambdaML is larger than that of \FP as \LambdaML has a higher communication cost. 
The relatively mild performance improvement of \FP with the increase of bandwidth suggests that \FP is more robust to different network settings. With 20X the bandwidth, compared with \LambdaML, \FP achieves comparable performances on \emph{ResNet101} and \emph{BERT-Large}, i.e., 12.2\% higher speed but 7.0\% higher cost when training \emph{ResNet101}, 12.9\% higher speed but 6.3\% higher cost when training \emph{BERT-Large}. The trade-offs in speed and cost are caused by the small differences in the tendencies of the policies of \FP and \LambdaML. When training \emph{AmoebaNet-D18/AmoebaNet-D36}, \FP improves the training speed by 6.8\%/14.0\% while reducing the cost by 6.4\%/38.6\%.
Such improvements are mostly attributed to \FP's optimized function memory allocation. 
This shows that even with the communication bottleneck removed, the memory allocation policy of \FP can still benefit serverless-based training, although by a smaller margin.

Despite the improvements that \FP achieves over existing serverless-based frameworks, a performance gap still exists between training with \FP and GPU-enabled VM instances due to the lack of GPU support in serverless function. We conduct preliminary comparison by training the models on a popular p3.2xlarge AWS instance (equipped with a V100 GPU). As none of the models can be trained on the single GPU without causing memory overflow, we adopt gradient accumulation to reduce the memory consumption. The micro-batch size used for gradient accumulation is 4, the same as the micro-batch size in \FP. The results reported in Fig.~\ref{fig:future_proof} show that GPU-based training can greatly outperform serverless CPU-based training in terms of cost, i.e. up to 90\% cost reduction. The cause is that the per data sample processing cost of a vCPU can be tens of times higher than that of a GPU. Fortunately, some of the serverless platforms, e.g. Alibaba Cloud, are recently equipping their serverless functions with GPU~\cite{ali_function_compute_instance_type}.
Similarly, we report the performance of training with a single serverless GPU function in Fig.~\ref{fig:future_proof}. Note that as GPU function has yet not been made fully available to users, we evaluate the training speed on a GPU of the same type as the GPU function and obtain the cost with the announced GPU function price. The results show that GPU function greatly narrows gap in the per data sample processing cost with VM GPU instance.
It is our next step of work to extend FuncPipe to such GPU function, evaluate its distributed training performance against VM GPUs and explore further optimization.

%% file: tex/related_work.tex
\section{Related Work}

\para{Pipeline in serverless-based training.} 
Dorylus~\cite{dorylus} is a pipelined framework with a hybrid structure, i.e., CPU servers with serverless functions, for training Graph Neural Network (GNN) models.
It exploits the inherent features of GNN to separate the computation tasks and uses serverless functions only for lightweight linear algebra operations. 
In contrast, our work exploits a serverless-based pipeline for training DNN models, which cannot be easily separated and trained the same way using Dorylus because they require much heavier computation and communication.
Hydrozoa~\cite{hydrozoa} proposes a pipelined framework that enables distributed training on GPU-enabled container instances. As serverless function has more stringent resource limits than container instance, our work focuses on providing more efficient communication design and careful co-optimization of model partition and resource allocation to tackle such resource challenges. Note that our optimization designs have the potential to benefit distributed training in other environments like the GPU-enabled container instances, as our preliminary experiments have demonstrated the benefits of our co-optimized model partition and resource allocation policy in GPU-based training. 

\para{Serverless communication.} 
Feng et al. propose two centralized storage-based methods for model synchronization~\cite{feng}.
However, such a design is generally of low efficiency due to the bandwidth bottleneck of the central nodes.
LambdaML proposes a more efficient decentralized scatter-reduce method, but it fails to utilize the available bandwidth fully~\cite{lambdaml}.
In parallel, other works focus on improving the performance of storage systems for higher communication efficiency. 
Pocket proposes a distributed data store that provides better elasticity and latency~\cite{pocket}.
Shredder designs a low-latency cloud store that supports in-storage computing~\cite{shredder}.
This line of work could be integrated with our pipelined storage-based communication approach to improving the network performance potentially.
Another choice is to use common \textit{NAT-traversal} techniques to enable direct communication among functions~\cite{nat1, nat3}. 
Direct communication can allow existing communication algorithms, e.g., ringAllreduce~\cite{ringallreduce}, to be used. 
However, NAT-traversal usually requires external servers that can cause communication bottlenecks.
The performance of using existing communication designs for serverless-based training with NAT-traversal remains unclear.

\para{Model partition and resource allocation in serverless.} 
Recent works have studied the model partition and resource allocation problem for serverless-based inference serving~\cite{gillis, amps}. 
These works aim at satisfying Service Level Objectives in latency while minimizing cost or further improving throughput. 
Gillis fixes the per-function memory allocation and optimizes model partition to lower inference cost with a reinforcement learning approach~\cite{gillis}. 
AMPS fixes the number of functions/partitions and co-optimizes the partition and memory allocation with a MIP formulation~\cite{amps}. 
Compared with inference, 
the optimization for distributed training, which is the focus of this work, includes more decision factors such as inner-stage data parallelism and synchronization cost and makes it more challenging to generate efficient model partition and resource configuration.


%% file: tex/conclusion.tex
\section{Conclusion}

In this paper, we presented the design and implementation of a novel pipelined serverless training framework called \FP. 
With the ever increasing interests in truly taking advantage of serverless computing, many researchers have looked at utilizing serverless functions to build scalable applications and improving serverless platforms~\cite{Wang2020-to,llama,Romero2021-ly,dorylus}.
Our key goal can be simply boiled down to understand \emph{how to allow DL practitioners to train models on serverless platforms} in a fast and low-cost manner, regardless of model size and training hyperparameters such as batch size that impact memory consumption. With three key designs---\1 the pipeline parallelism for model partitions, \2 the communication-efficient scatter-reduce, and \3 the co-optimization of partition and resource allocation policy, \FP was able to overcome the memory and bandwidth limitations of serverless platforms. 
We demonstrated the benefits of \FP, i.e. \textbf{1.3X-2.2X} training speedup and \textbf{7\%-77\%} cost reduction compared to state-of-the-art serverless training frameworks~\cite{lambdaml}, by testing with four commonly used models and on two popular serverless providers in numerous settings. Interestingly, we observed that the benefits of \FP remain even if the bandwidth of serverless functions increases to a level comparable to today's VM bandwidth. This observation suggests the relevance of \FP techniques even as the cloud providers continue to improve serverless infrastructure.


%% file: tex/ack.tex
\begin{acks}
This work was supported in part by the National Natural Science Foundation of China (No. 42050105, 62072302, 62020106005, 62061146002, 61960206002), the Program of Shanghai Academic/Technology Research Leader under Grant No. 18XD1401800, the US National Science Foundation under Grant NGSDI-2105564, and VMWare. We thank our shepherd Michael
Ferdman, and the anonymous reviewers.
\end{acks}


%% file: tex/notation.tex

\section{Notations in \S \ref{sec:design_coopt}}\label{app:notation}


 \begin{table*}[h]
   \centering
    \small
 \begin{tabular}{ll} 
     \toprule 
     \textbf{Notation} & \textbf{Definition} \\
     \midrule
 $s_{0}$& basic memory consumption of a serverless worker\\
 $M$& total number of micro-batches\\
 $L$&  number of model layers. \\
 $P$& unit price of serverless function \\
 $t_{lat}$& latency from serverless worker to cloud storage\\
 $s_{i}$& model size of layer $i$ \\
 $a_i$& size of activations of layer $i$ per micro-batch\\ 
 $o_i$& size of output of layer $i$ per micro-batch \\
 $g_i$& size of gradients from layer $i$ to layer $i-1$ per micro-batch \\
 $\beta$& slowdown factor for computation due to resource contention\\
 \midrule
 $K$& number of data parallelism options\\
 $D_k$& value of $k$-th data parallelism option\\
 $J$& number of resource allocation options\\
 $M_j$& memory size of $j$-th resource  option \\ 
 $W_j$& bandwidth of $j$-th resource  option \\
 $T_{fc}^{i,j}$& forward computation time of layer $i$ with $j$-th resource option \\
 $T_{bc}^{i,j}$& backward computation time of layer $i$ with $j$-th resource option \\
 \midrule
 $x_{i}$& $\{0,1\}$, $1$ means model is partitioned between layers $i$ and $i+1$ \\
 $y_{k}$& $\{0,1\}$, $1$ means the $k$-th data parallelism option $D_j$ is chosen \\
 $z_{i,j}$& $\{0,1\}$, $1$ means layer $i$ workers have $j$-th memory size $M_j$ \\
 \midrule
 $t_{iter}$& iteration time \\
 $c_{iter}$& iteration cost \\
 $t_{f}$&  forward time for full forward pipeline\\
 $t_{f}^0$&  time for one micro-batch traverse forward pipeline\\
 $\Delta_f$ & lag between micro-batches at end of forward pipeline\\
 $t_{b}^{i}$& backward time until layer $i$ completes computation\\
 $t_{s}^{i}$ & model synchronizing time at layer $i$ \\
 $t_{fu}^i$& time for layer $i$ to upload its output to storage. \\
 $t_{fd}^i$& time for layer $i+1$ to download input from storage. \\
 $t_{bu}^i$& time for $i$ to upload gradient output to storage. \\
 $t_{bd}^i$& time for layer $i-1$ to be download gradient from storage. \\
 $d$ & degree of data parallelism, $d = \sum_{i=1}^K y_k D_k$\\
 $\mu$ & number of micro-batches per worker, $\mu = M/d$\\
 $m_i$ & memory size of layer $i$ worker, $m_i = \sum_{j=1}^J z_{i,j} M_j$\\
 $w_i$ & bandwidth of layer $i$ worker, $w_i = \sum_{j=1}^J z_{i,j} W_j$\\
 $\hat a_i$& accumulated activation size at layer $i$ (accumulated forwardly).\\
 $\hat s_i$& accumulated model size at layer $i$ (accumulated forwardly).\\
 $\tilde s_i$& accumulated model size at layer $i$ (accumulated backwardly).\\
 $\hat t_{fc}^{i}$& accumulated forward computation time at layer $i$ (accumulated forwardly).\\
 $\tilde t_{bc}^i$& accumulated backward computation time at layer $i$ (accumulated backwardly).\\
 \bottomrule
 \end{tabular}
 \caption{Notations in \S \ref{sec:design_coopt}}
  \label{table:notation}
 \end{table*}

%% file: tex/model_app.tex
\section{Backward Time}\label{app:model}

The backward computation time $t_{bc}^i$ of layer $i$, the upload and download time  $t_{bu}^i$, $t_{bd}^i$ between layers $i$ and $i-1$ are given by
\[
\begin{aligned}
t_{bc}^{i} &= \beta  \sum_{j=1}^{J} z_{i,j}T_{bc}^{i,j}, & 1\le i\le L,\\
t_{bu}^{i} &= x_{i-1}\left(\sum_{j=1}^J z_{i,j} \frac{g_i}{W_j}+ t_{lat}\right), & 2\le i\le L, \\
t_{bd}^{i} &= x_{i-1}\left(\sum_{j=1}^J z_{(i-1),j}\frac{g_{i}}{W_j}+ t_{lat}\right), & 2\le i\le L,
\end{aligned}
\]
where $g_i$ is the gradient size from layer $i$ to layer $i-1$. We introduce a tilde operator  similar to the hat operator in \eqref{eq:forward_cs}, except that it accumulates the quantities backwardly. The cumulative backward computation time $\tilde t_{bc}^i$ from the previous partition boundary down to layer $i$ is given by
\begin{equation}\label{eq:backward_cs}
\tilde t_{bc}^{L} = t_{bc}^{L}, \quad \tilde t_{bc}^i = t_{bc}^{i} + \tilde t_{bc}^{i+1} (1-x_i), \quad 1\le i\le L-1.
\end{equation}

For each $1\le i\le L$, define
\begin{equation}\label{eq:backward_time}
t_b^i = \sum_{k=i}^{L}t_{bc}^{k} + \sum_{k=i+1}^{L}  (t_{bu}^{k} + t_{bd}^{k})  + (\mu-1) \Delta_b^i,
\end{equation}
where
\[
\Delta_b^i = \max \left\{\tilde t_{bc}^{i:L},    t_{bu}^{(i+1):L} ,   t_{bd}^{(i+1):L} \right\}.
\]
When $i$ is the lowest layer of a partition, $t_b^i$ is the computation completion time of that partition, and $\Delta_b^i$ is the corresponding lag between consecutive micro-batches. Note that $t_b^i \ge t_b^{i'}$ if $i'\ge i$ and layers $i$ and $i'$ belong to the same partition.

%% file: tex/linearization.tex
\section{Linearization}
\label{sec:linearization}

First we present the major linearization techniques used to convert the non-linear binary integer programming to MIQP:

\para{Technique 1:} Linearizing the multiplication of two binary variables. $x,y \in \{0,1\}$, $xy$ can be linearized as follows:
\begin{equation*}
    \begin{aligned}
        f = xy\\
        f\leq x  \\ 
        f\leq y  \\
        f\geq x + y -1\\
        f \in \{0,1\}
    \end{aligned}
\end{equation*}

\para{Technique 2:} Linearizing the multiplication of a continuous variable and a binary variable. $x \in \{0,1\}$, $y\in [a,b]$ is a continuous variable, $xy$ can be linearized as follows:
\begin{equation*}
    \begin{aligned}
        f = xy\\
        f\leq y  \\ 
        f\geq y - b(1-x)\\
        ax \leq y \leq bx
    \end{aligned}
\end{equation*}

\para{Technique 3:} Linearizing of the $\max$ operator. $x,y,z$ are continuous variables, $\max\{x,y,z\}$ can be linearized as follows:
\begin{equation*}
    \begin{aligned}
        f = \max\{x,y,z\}\\
        x\leq f, y\leq f, z\leq f\\
        x\geq f - H(1-l_1)\\
        y\geq f - H(1-l_2) \\
        z\geq f - H(1-l_3) \\
        l_1 + l_2 + l_3 \geq 1\\
        l_1, l_2, l_3 \in \{0,1\}
    \end{aligned}
\end{equation*}
where $H$ is a large constant. Next we introduce how we linearize the formulation in detail. 
\begin{enumerate}
    \item \textit{Linearizing the equality constraint for the cumulative values $\hat t_{fc}^{i}$, $\tilde t_{bc}^{i}$, $\hat s_{i}$, $\tilde s_{i}$ and $\hat a_{i}$.} We introduce $\hat t_{fc}^{i}$, $\tilde t_{bc}^{i}$, $\hat s_{i}$, $\tilde s_{i}$ and $\hat a_{i}$ as continuous variables and linearize their equality constraints. We use $\tilde t_{bc}^{i}$ in \eqref{eq:backward_cs} as an example and it is similar with the others. We can write $\tilde t_{bc}^{i}$ as:
    \begin{equation*}
        \begin{aligned}
            r_{i} &= 1 - x_i \\
            \tilde t_{bc}^{i} &= t_{bc}^{i} + \tilde t_{bc}^{i+1}r_{i} \\
            & = \sum_{q=i}^{L}t_{bc}^{q}\prod_{p=i}^{q-1}r_{p}
        \end{aligned}
    \end{equation*}
    Since $r_i$ is a binary variable, $\prod_{p=i}^{q-1}r_{p}$ can be converted to a new binary variable $\dot r_{i,q}$ by recursively performing linearization with \textbf{Technique 1}. Then continuous variable $\tilde t_{bc}^{i}$ satisfies the following constraint
    \begin{equation*}
        \begin{aligned}
            \tilde t_{bc}^{i} &= \sum_{q=i}^{L}t_{bc}^{i} \dot r_{i,q} \\
            &= \beta  \sum_{q=i}^{L}\sum_{j=1}^{J} z_{i,j}\dot r_{i,q}T_{bc}^{i,j}
        \end{aligned}
    \end{equation*}
    $z_{i,j}$ and $\dot r_{i,q}$ are both binary variables, thus $z_{i,j}\dot r_{i,q}$ can be linearized applying \textbf{Technique 1}.
    
    \item \textit{Linearizing the equality constraint for $t_{fu}^{i}$, $t_{fd}^{i}$, $t_{bu}^{i}$ and $t_{bd}^{i}$.} We introduce $t_{fu}^{i}$, $t_{fd}^{i}$, $t_{bu}^{i}$ and $t_{bd}^{i}$ as continuous variables and linearize their equality constraints. We use $t_{fu}^{i}$ as an example and it is similar with the others. We can write $t_{fu}^{i}$ in \eqref{eq:forward_layer2} as:
    \begin{equation*}
        \label{eq:linear_fu}
        \begin{aligned}
            t_{fu}^{i} &= \sum_{j=1}^J x_iz_{i,j} \frac{o_i}{W_j}+ x_it_{lat}
        \end{aligned}
    \end{equation*}
    $x_i$ and $z_{i,k}$ are both binary variables, thus $x_iz_{i,j}$ can be linearized applying \textbf{Technique 1}.
    
    \item \textit{Linearizing forward time $t_{f}$ and backward time $t_{b}^{i}$.} We use $t_{b}^{i}$ as an example and it is similar with $t_{f}$. Linearizing $t_{b}^{i}$ in \eqref{eq:backward_time} is equal to linearizing $(\mu-1) \Delta_b^i$. Since $\Delta_b^i$ is the $\max$ of a set of continuous variables, it can be presented as a continuous variable with linear constraints using \textbf{Technique 3}. Expand $(\mu-1) \Delta_b^i$, we have
    \begin{equation*}
    \begin{aligned}
    (\mu-1) \Delta_b^i = \sum_{k=1}^K \Delta_b^iy_k \frac{M}{D_k} - \Delta_b^i
    \end{aligned}
    \end{equation*}
    Since $\Delta_b^i$ is a continuous variable and $y_k$ is a binary variable, $\Delta_b^iy_k$ can be linearized applying \textbf{Technique 2}.
    
    \item \textit{Linearizing $t_{s}^{i}$.} Expand \eqref{eq:sync_time}, we have
    \begin{equation*}
    \begin{aligned}
    t_{s}^{i} = \sum_{j=1}^J (z_{ij}\tilde s_{i} - y_{1}z_{ij}\tilde s_{i}) \frac{\gamma}{W_j} + (1 - y_{1})t_{lat} \cdot \delta,
    \end{aligned}
    \end{equation*}
    $z_{ij}$ and $y_{1}$ are binary variables, $\tilde s_{i}$ is a continuous variable, thus we can first linearize $z_{ij}\tilde s_{i}$ using \textbf{Technique 2} and then further linearize $y_{1}z_{ij}\tilde s_{i}$ by applying \textbf{Technique 2} again.
    
    \item \textit{Linearizing full iteration time $t_{iter}$.} So far we have linearized $t_f$, $t_b^{i}$ and $t_s^{i}$ in \eqref{eq:iter_time}. We can further remove the $\max$ operator using \textbf{Technique 3}. 
    
    \item \textit{Linearizing total memory allocation $c_{mem}$.} Expand \eqref{eq:mem_cost}, we have
    \begin{equation*}
    \begin{aligned}
    c_{mem} = \sum_{i=1}^{L-1}\sum_{k=1}^K \sum_{j=1}^J x_{i} y_k z_{i,j} D_k M_j + \sum_{k=1}^K \sum_{j=1}^J y_k z_{L,j} D_k M_j 
    \end{aligned}
    \end{equation*}
    Since $x_{i}$, $y_k$, and $z_{i,j}$ are all binary variables, $x_{i} y_k z_{i,j}$ and $y_k z_{L,j}$ can be linearized using \textbf{Technique 1}.
    
    \item \textit{Linearizing memory constraint.} At last, we linearize the memory constraint, the first constraint in \eqref{prob:opt}. Expand the constraint, we have
    \begin{equation*}
    \begin{aligned}
    \sum_{k=1}^K \hat a_{i} y_k \frac{M}{D_k} +  4\hat s_i -2 \hat s_i y_{1}  + s_0 \leq \sum_{j=1}^J z_{i,j} M_j
    \end{aligned}
    \end{equation*}
    
    $\hat a_{i} y_k$ and $\hat s_i y_{1}$ can both be linearized with \textbf{Technique 2}. 
\end{enumerate}

After linearization, $t_{iter}$, $c_{mem}$ and the constraints in \eqref{prob:opt} are all in linear form. $c_{iter}$ (\eqref{eq:money_cost}) and the objective function are quadratic. The formulation becomes a mixed-integer quadratic program. It has a total of $\max\{o(JL^2), o(JKL)\}$ integer variables, $\max\{o(JL),o(KL)\}$ continuous variables and $\max\{o(JL^2), o(JKL)\}$ linear constraints.

%% file: tex/api_app.tex
\section{A FuncPipe function example}\label{app:api_example}
Below is a code example for training with \FP. As highlighted in orange, only minimal changes to the Pytorch training code are required.

\vspace{.5em}
\includegraphics[width=.7\linewidth]{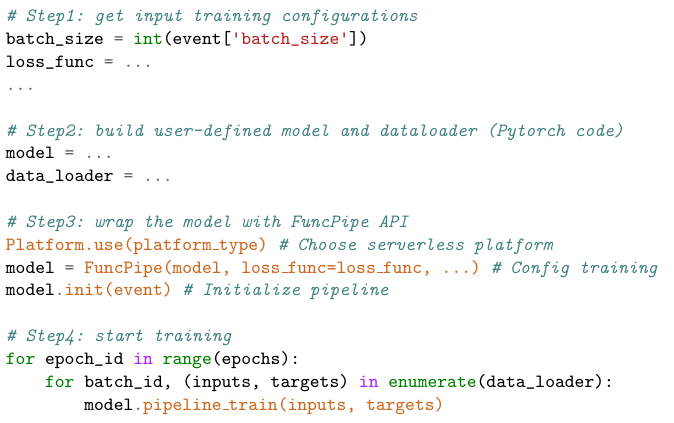}


%% file: tex/eval_model_accuracy.tex
\section{Performance Model Accuracy}\label{app:model_accuracy}





Table~\ref{tab:pred_error} displays the prediction error in training time for the measured points of \FP in Fig.~\ref{fig:overall_perf}. The results show that our performance model achieves an average prediction error of less than 12\%.
The largest error happens when training \emph{Amoebanet-D36} with a global batch size of $256$.
We note that this error is mainly caused by the unexpected bandwidth variation; other model training is less impacted as they use fewer serverless workers and are less subject to the performance interference among workers. 
We leave the consideration of such interference in our performance model as part of future work.

\begin{table}[h]
    \centering
    \fontsize{8}{10}\selectfont  
\begin{tabular}{l|ccc|c}
    \toprule
\diagbox[width=10em,trim=l]{\textbf{Model}}{\textbf{Batchsize}} & 16 & 64 & 256  & Average  \\
\hline
\emph{ResNet101} & 5.9\% & 11.2\% & 15.4\% & 10.8\%  \\
\emph{Amoebanet-D18} & 13.3\% & 9.0\% & 10.6\% & 11.0\%  \\
\emph{Amoebanet-D36} & 10.8\% & 4.0\% & 18.1\% & 11.0\%  \\
\emph{Bert-large} & 9.8\% & 11.0\% & 16.4\% & 12.4\%  \\
\hline
Average & 9.9\%  & 8.8\%  & 15.1\%  & 11.3\% \\
    \bottomrule
\end{tabular}\vspace{0cm}
    \caption{\textbf{Prediction error of \FP training tasks.}
     Our performance model achieves an average prediction error of less than 12\%.
    }
    \label{tab:pred_error}
\end{table}


%% file: main.bbl

\begin{thebibliography}{77}


\ifx \showCODEN    \undefined \def \showCODEN     #1{\unskip}     \fi
\ifx \showDOI      \undefined \def \showDOI       #1{#1}\fi
\ifx \showISBNx    \undefined \def \showISBNx     #1{\unskip}     \fi
\ifx \showISBNxiii \undefined \def \showISBNxiii  #1{\unskip}     \fi
\ifx \showISSN     \undefined \def \showISSN      #1{\unskip}     \fi
\ifx \showLCCN     \undefined \def \showLCCN      #1{\unskip}     \fi
\ifx \shownote     \undefined \def \shownote      #1{#1}          \fi
\ifx \showarticletitle \undefined \def \showarticletitle #1{#1}   \fi
\ifx \showURL      \undefined \def \showURL       {\relax}        \fi
\providecommand\bibfield[2]{#2}
\providecommand\bibinfo[2]{#2}
\providecommand\natexlab[1]{#1}
\providecommand\showeprint[2][]{arXiv:#2}

\bibitem[Akkus et~al\mbox{.}(2018)]%
        {serverless_application2}
\bibfield{author}{\bibinfo{person}{Istemi~Ekin Akkus},
  \bibinfo{person}{Ruichuan Chen}, \bibinfo{person}{Ivica Rimac},
  \bibinfo{person}{Manuel Stein}, \bibinfo{person}{Klaus Satzke},
  \bibinfo{person}{Andre Beck}, \bibinfo{person}{Paarijaat Aditya}, {and}
  \bibinfo{person}{Volker Hilt}.} \bibinfo{year}{2018}\natexlab{}.
\newblock \showarticletitle{{SAND: Towards High-Performance Serverless
  Computing}}. In \bibinfo{booktitle}{\emph{2018 Usenix Annual Technical
  Conference (USENIX ATC 18)}}. \bibinfo{pages}{923--935}.
\newblock


\bibitem[Alibaba(2022a)]%
        {ali_function_compute}
\bibfield{author}{\bibinfo{person}{Alibaba}.} \bibinfo{year}{2022}\natexlab{a}.
\newblock \bibinfo{title}{Alibaba Cloud Function Compute}.
\newblock
\newblock
\newblock
\shownote{\url{https://www.aliyun.com/product/fc}}.


\bibitem[Alibaba(2022b)]%
        {ali_function_compute_instance_type}
\bibfield{author}{\bibinfo{person}{Alibaba}.} \bibinfo{year}{2022}\natexlab{b}.
\newblock \bibinfo{title}{Alibaba Cloud Function Compute Instance Type}.
\newblock
\newblock
\newblock
\shownote{\url{https://help.aliyun.com/document_detail/179379.html}}.


\bibitem[Alibaba(2022c)]%
        {ali_oss}
\bibfield{author}{\bibinfo{person}{Alibaba}.} \bibinfo{year}{2022}\natexlab{c}.
\newblock \bibinfo{title}{Alibaba Cloud Object Storage Service}.
\newblock
\newblock
\newblock
\shownote{\url{https://www.aliyun.com/product/oss}}.


\bibitem[Alipourfard et~al\mbox{.}(2017)]%
        {cherrypick}
\bibfield{author}{\bibinfo{person}{Omid Alipourfard},
  \bibinfo{person}{Hongqiang~Harry Liu}, \bibinfo{person}{Jianshu Chen},
  \bibinfo{person}{Shivaram Venkataraman}, \bibinfo{person}{Minlan Yu}, {and}
  \bibinfo{person}{Ming Zhang}.} \bibinfo{year}{2017}\natexlab{}.
\newblock \showarticletitle{{CherryPick}: Adaptively Unearthing the Best Cloud
  Configurations for Big Data Analytics}. In \bibinfo{booktitle}{\emph{14th
  USENIX Symposium on Networked Systems Design and Implementation (NSDI 17)}}.
  \bibinfo{pages}{469--482}.
\newblock


\bibitem[Amazon(2022)]%
        {aws_lambda}
\bibfield{author}{\bibinfo{person}{Amazon}.} \bibinfo{year}{2022}\natexlab{}.
\newblock \bibinfo{title}{AWS Lambda}.
\newblock
\newblock
\newblock
\shownote{\url{https://www.aliyun.com/product/fc}}.


\bibitem[Awan et~al\mbox{.}(2020)]%
        {hybrid_parallelism2}
\bibfield{author}{\bibinfo{person}{Ammar~Ahmad Awan}, \bibinfo{person}{Arpan
  Jain}, \bibinfo{person}{Quentin Anthony}, \bibinfo{person}{Hari Subramoni},
  {and} \bibinfo{person}{Dhabaleswar~K Panda}.}
  \bibinfo{year}{2020}\natexlab{}.
\newblock \showarticletitle{HyPar-Flow: exploiting MPI and Keras for scalable
  hybrid-parallel DNN training with tensorflow}. In
  \bibinfo{booktitle}{\emph{International Conference on High Performance
  Computing}}. \bibinfo{pages}{83--103}.
\newblock


\bibitem[Bian et~al\mbox{.}(2021)]%
        {model_parallelism4}
\bibfield{author}{\bibinfo{person}{Zhengda Bian}, \bibinfo{person}{Qifan Xu},
  \bibinfo{person}{Boxiang Wang}, {and} \bibinfo{person}{Yang You}.}
  \bibinfo{year}{2021}\natexlab{}.
\newblock \showarticletitle{Maximizing Parallelism in Distributed Training for
  Huge Neural Networks}.
\newblock \bibinfo{journal}{\emph{arXiv preprint arXiv:2105.14450}}
  (\bibinfo{year}{2021}).
\newblock


\bibitem[Bojja~Venkatakrishnan et~al\mbox{.}(2019)]%
        {nips_partition_model}
\bibfield{author}{\bibinfo{person}{Shaileshh Bojja~Venkatakrishnan},
  \bibinfo{person}{Shreyan Gupta}, \bibinfo{person}{Hongzi Mao},
  \bibinfo{person}{Mohammad Alizadeh}, {et~al\mbox{.}}}
  \bibinfo{year}{2019}\natexlab{}.
\newblock \showarticletitle{Learning Generalizable Device Placement Algorithms
  for Distributed Machine Learning}.
\newblock \bibinfo{journal}{\emph{Advances in Neural Information Processing
  Systems}}  \bibinfo{volume}{32} (\bibinfo{year}{2019}).
\newblock


\bibitem[Brochu et~al\mbox{.}(2010)]%
        {bayes}
\bibfield{author}{\bibinfo{person}{Eric Brochu}, \bibinfo{person}{Vlad~M Cora},
  {and} \bibinfo{person}{Nando De~Freitas}.} \bibinfo{year}{2010}\natexlab{}.
\newblock \showarticletitle{A tutorial on Bayesian optimization of expensive
  cost functions, with application to active user modeling and hierarchical
  reinforcement learning}.
\newblock \bibinfo{journal}{\emph{arXiv preprint arXiv:1012.2599}}
  (\bibinfo{year}{2010}).
\newblock


\bibitem[Carreira et~al\mbox{.}(2019)]%
        {cirrus}
\bibfield{author}{\bibinfo{person}{Joao Carreira}, \bibinfo{person}{Pedro
  Fonseca}, \bibinfo{person}{Alexey Tumanov}, \bibinfo{person}{Andrew Zhang},
  {and} \bibinfo{person}{Randy Katz}.} \bibinfo{year}{2019}\natexlab{}.
\newblock \showarticletitle{Cirrus: A serverless framework for end-to-end ml
  workflows}. In \bibinfo{booktitle}{\emph{Proceedings of the ACM Symposium on
  Cloud Computing}}. \bibinfo{pages}{13--24}.
\newblock


\bibitem[Chen et~al\mbox{.}(2018)]%
        {model_parallelism2}
\bibfield{author}{\bibinfo{person}{Chi-Chung Chen}, \bibinfo{person}{Chia-Lin
  Yang}, {and} \bibinfo{person}{Hsiang-Yun Cheng}.}
  \bibinfo{year}{2018}\natexlab{}.
\newblock \showarticletitle{Efficient and robust parallel dnn training through
  model parallelism on multi-gpu platform}.
\newblock \bibinfo{journal}{\emph{arXiv preprint arXiv:1809.02839}}
  (\bibinfo{year}{2018}).
\newblock


\bibitem[Chu et~al\mbox{.}(2020)]%
        {distributed_nvlink}
\bibfield{author}{\bibinfo{person}{Ching-Hsiang Chu}, \bibinfo{person}{Pouya
  Kousha}, \bibinfo{person}{Ammar~Ahmad Awan}, \bibinfo{person}{Kawthar~Shafie
  Khorassani}, \bibinfo{person}{Hari Subramoni}, {and}
  \bibinfo{person}{Dhabaleswar~K Panda}.} \bibinfo{year}{2020}\natexlab{}.
\newblock \showarticletitle{Nv-group: link-efficient reduction for distributed
  deep learning on modern dense gpu systems}. In
  \bibinfo{booktitle}{\emph{Proceedings of the 34th ACM International
  Conference on Supercomputing}}. \bibinfo{pages}{1--12}.
\newblock


\bibitem[Devarakonda et~al\mbox{.}(2017)]%
        {ga3}
\bibfield{author}{\bibinfo{person}{Aditya Devarakonda}, \bibinfo{person}{Maxim
  Naumov}, {and} \bibinfo{person}{Michael Garland}.}
  \bibinfo{year}{2017}\natexlab{}.
\newblock \showarticletitle{Adabatch: Adaptive batch sizes for training deep
  neural networks}.
\newblock \bibinfo{journal}{\emph{arXiv preprint arXiv:1712.02029}}
  (\bibinfo{year}{2017}).
\newblock


\bibitem[Domhan et~al\mbox{.}(2015)]%
        {hyper-tuning-ns-2}
\bibfield{author}{\bibinfo{person}{Tobias Domhan}, \bibinfo{person}{Jost~Tobias
  Springenberg}, {and} \bibinfo{person}{Frank Hutter}.}
  \bibinfo{year}{2015}\natexlab{}.
\newblock \showarticletitle{Speeding up automatic hyperparameter optimization
  of deep neural networks by extrapolation of learning curves}. In
  \bibinfo{booktitle}{\emph{Twenty-fourth international joint conference on
  artificial intelligence}}.
\newblock


\bibitem[Eismann et~al\mbox{.}(2020)]%
        {serverless_application4}
\bibfield{author}{\bibinfo{person}{Simon Eismann}, \bibinfo{person}{Joel
  Scheuner}, \bibinfo{person}{Erwin Van~Eyk}, \bibinfo{person}{Maximilian
  Schwinger}, \bibinfo{person}{Johannes Grohmann}, \bibinfo{person}{Nikolas
  Herbst}, \bibinfo{person}{Cristina~L Abad}, {and} \bibinfo{person}{Alexandru
  Iosup}.} \bibinfo{year}{2020}\natexlab{}.
\newblock \showarticletitle{Serverless applications: Why, when, and how?}
\newblock \bibinfo{journal}{\emph{IEEE Software}} \bibinfo{volume}{38},
  \bibinfo{number}{1} (\bibinfo{year}{2020}), \bibinfo{pages}{32--39}.
\newblock


\bibitem[Fan et~al\mbox{.}(2021)]%
        {dapple}
\bibfield{author}{\bibinfo{person}{Shiqing Fan}, \bibinfo{person}{Yi Rong},
  \bibinfo{person}{Chen Meng}, \bibinfo{person}{Zongyan Cao},
  \bibinfo{person}{Siyu Wang}, \bibinfo{person}{Zhen Zheng},
  \bibinfo{person}{Chuan Wu}, \bibinfo{person}{Guoping Long},
  \bibinfo{person}{Jun Yang}, \bibinfo{person}{Lixue Xia}, {et~al\mbox{.}}}
  \bibinfo{year}{2021}\natexlab{}.
\newblock \showarticletitle{DAPPLE: A pipelined data parallel approach for
  training large models}. In \bibinfo{booktitle}{\emph{Proceedings of the 26th
  ACM SIGPLAN Symposium on Principles and Practice of Parallel Programming}}.
  \bibinfo{pages}{431--445}.
\newblock


\bibitem[Feng et~al\mbox{.}(2018)]%
        {feng}
\bibfield{author}{\bibinfo{person}{Lang Feng}, \bibinfo{person}{Prabhakar
  Kudva}, \bibinfo{person}{Dilma Da~Silva}, {and} \bibinfo{person}{Jiang Hu}.}
  \bibinfo{year}{2018}\natexlab{}.
\newblock \showarticletitle{Exploring serverless computing for neural network
  training}. In \bibinfo{booktitle}{\emph{2018 IEEE 11th international
  conference on cloud computing (CLOUD)}}. IEEE, \bibinfo{pages}{334--341}.
\newblock


\bibitem[Fouladi et~al\mbox{.}(2019)]%
        {nat1}
\bibfield{author}{\bibinfo{person}{Sadjad Fouladi}, \bibinfo{person}{Francisco
  Romero}, \bibinfo{person}{Dan Iter}, \bibinfo{person}{Qian Li},
  \bibinfo{person}{Shuvo Chatterjee}, \bibinfo{person}{Christos Kozyrakis},
  \bibinfo{person}{Matei Zaharia}, {and} \bibinfo{person}{Keith Winstein}.}
  \bibinfo{year}{2019}\natexlab{}.
\newblock \showarticletitle{From laptop to lambda: Outsourcing everyday jobs to
  thousands of transient functional containers}. In
  \bibinfo{booktitle}{\emph{2019 USENIX Annual Technical Conference (USENIX ATC
  19)}}. \bibinfo{pages}{475--488}.
\newblock


\bibitem[Geng et~al\mbox{.}(2019)]%
        {hybrid_parallelism3}
\bibfield{author}{\bibinfo{person}{Jinkun Geng}, \bibinfo{person}{Dan Li},
  {and} \bibinfo{person}{Shuai Wang}.} \bibinfo{year}{2019}\natexlab{}.
\newblock \showarticletitle{Horizontal or vertical? a hybrid approach to
  large-scale distributed machine learning}. In
  \bibinfo{booktitle}{\emph{Proceedings of the 10th Workshop on Scientific
  Cloud Computing}}. \bibinfo{pages}{1--4}.
\newblock


\bibitem[Guo et~al\mbox{.}(2022)]%
        {hydrozoa}
\bibfield{author}{\bibinfo{person}{Runsheng Guo}, \bibinfo{person}{Victor Guo},
  \bibinfo{person}{Antonio Kim}, \bibinfo{person}{Josh Hildred}, {and}
  \bibinfo{person}{Khuzaima Daudjee}.} \bibinfo{year}{2022}\natexlab{}.
\newblock \showarticletitle{Hydrozoa: Dynamic Hybrid-Parallel DNN Training on
  Serverless Containers}.
\newblock \bibinfo{journal}{\emph{Proceedings of Machine Learning and Systems}}
   \bibinfo{volume}{4} (\bibinfo{year}{2022}), \bibinfo{pages}{779--794}.
\newblock


\bibitem[Gurobi(2022a)]%
        {gurobi}
\bibfield{author}{\bibinfo{person}{Gurobi}.} \bibinfo{year}{2022}\natexlab{a}.
\newblock \bibinfo{title}{Gurobi - The Fastest Solver}.
\newblock
\newblock
\newblock
\shownote{\url{https://www.gurobi.com}}.


\bibitem[Gurobi(2022b)]%
        {floating_license}
\bibfield{author}{\bibinfo{person}{Gurobi}.} \bibinfo{year}{2022}\natexlab{b}.
\newblock \bibinfo{title}{Setting up and using a Floating license}.
\newblock
\newblock
\newblock
\shownote{\url{https://www.gurobi.com/documentation/9.5/quickstart_mac/setting_up_and_using_a_flo.html}}.


\bibitem[Hafeez et~al\mbox{.}(2021)]%
        {midware_partition_model}
\bibfield{author}{\bibinfo{person}{Ubaid~Ullah Hafeez}, \bibinfo{person}{Xiao
  Sun}, \bibinfo{person}{Anshul Gandhi}, {and} \bibinfo{person}{Zhenhua Liu}.}
  \bibinfo{year}{2021}\natexlab{}.
\newblock \showarticletitle{Towards optimal placement and scheduling of DNN
  operations with Pesto}. In \bibinfo{booktitle}{\emph{Proceedings of the 22nd
  International Middleware Conference}}. \bibinfo{pages}{39--51}.
\newblock


\bibitem[Hendrickson et~al\mbox{.}(2016)]%
        {serverless_application1}
\bibfield{author}{\bibinfo{person}{Scott Hendrickson}, \bibinfo{person}{Stephen
  Sturdevant}, \bibinfo{person}{Tyler Harter}, {and}
  \bibinfo{person}{Venkataramani.}} \bibinfo{year}{2016}\natexlab{}.
\newblock \showarticletitle{Serverless Computation with {OpenLambda}}. In
  \bibinfo{booktitle}{\emph{8th USENIX Workshop on Hot Topics in Cloud
  Computing (HotCloud 16)}}.
\newblock


\bibitem[Hu et~al\mbox{.}(2021)]%
        {helios}
\bibfield{author}{\bibinfo{person}{Qinghao Hu}, \bibinfo{person}{Peng Sun},
  \bibinfo{person}{Shengen Yan}, \bibinfo{person}{Yonggang Wen}, {and}
  \bibinfo{person}{Tianwei Zhang}.} \bibinfo{year}{2021}\natexlab{}.
\newblock \showarticletitle{Characterization and prediction of deep learning
  workloads in large-scale gpu datacenters}. In
  \bibinfo{booktitle}{\emph{Proceedings of the International Conference for
  High Performance Computing, Networking, Storage and Analysis}}.
  \bibinfo{pages}{1--15}.
\newblock


\bibitem[Huang et~al\mbox{.}(2019)]%
        {gpipe}
\bibfield{author}{\bibinfo{person}{Yanping Huang}, \bibinfo{person}{Youlong
  Cheng}, \bibinfo{person}{Ankur Bapna}, \bibinfo{person}{Orhan Firat},
  \bibinfo{person}{Dehao Chen}, \bibinfo{person}{Mia Chen},
  \bibinfo{person}{HyoukJoong Lee}, \bibinfo{person}{Jiquan Ngiam},
  \bibinfo{person}{Quoc~V Le}, \bibinfo{person}{Yonghui Wu}, {et~al\mbox{.}}}
  \bibinfo{year}{2019}\natexlab{}.
\newblock \showarticletitle{Gpipe: Efficient training of giant neural networks
  using pipeline parallelism}.
\newblock \bibinfo{journal}{\emph{Advances in neural information processing
  systems}}  \bibinfo{volume}{32} (\bibinfo{year}{2019}).
\newblock


\bibitem[Jain et~al\mbox{.}(2020a)]%
        {model_parallelism3}
\bibfield{author}{\bibinfo{person}{Arpan Jain}, \bibinfo{person}{Ammar~Ahmad
  Awan}, \bibinfo{person}{Asmaa~M Aljuhani}, \bibinfo{person}{Jahanzeb~Maqbool
  Hashmi}, \bibinfo{person}{Quentin~G Anthony}, \bibinfo{person}{Hari
  Subramoni}, \bibinfo{person}{Dhableswar~K Panda}, \bibinfo{person}{Raghu
  Machiraju}, {and} \bibinfo{person}{Anil Parwani}.}
  \bibinfo{year}{2020}\natexlab{a}.
\newblock \showarticletitle{Gems: Gpu-enabled memory-aware model-parallelism
  system for distributed dnn training}. In \bibinfo{booktitle}{\emph{SC20:
  International Conference for High Performance Computing, Networking, Storage
  and Analysis}}. IEEE, \bibinfo{pages}{1--15}.
\newblock


\bibitem[Jain et~al\mbox{.}(2020b)]%
        {gems}
\bibfield{author}{\bibinfo{person}{Arpan Jain}, \bibinfo{person}{Ammar~Ahmad
  Awan}, \bibinfo{person}{Asmaa~M Aljuhani}, \bibinfo{person}{Jahanzeb~Maqbool
  Hashmi}, \bibinfo{person}{Quentin~G Anthony}, \bibinfo{person}{Hari
  Subramoni}, \bibinfo{person}{Dhableswar~K Panda}, \bibinfo{person}{Raghu
  Machiraju}, {and} \bibinfo{person}{Anil Parwani}.}
  \bibinfo{year}{2020}\natexlab{b}.
\newblock \showarticletitle{Gems: Gpu-enabled memory-aware model-parallelism
  system for distributed dnn training}. In \bibinfo{booktitle}{\emph{SC20:
  International Conference for High Performance Computing, Networking, Storage
  and Analysis}}. IEEE, \bibinfo{pages}{1--15}.
\newblock


\bibitem[Jarachanthan et~al\mbox{.}(2021)]%
        {amps}
\bibfield{author}{\bibinfo{person}{Jananie Jarachanthan}, \bibinfo{person}{Li
  Chen}, \bibinfo{person}{Fei Xu}, {and} \bibinfo{person}{Bo Li}.}
  \bibinfo{year}{2021}\natexlab{}.
\newblock \showarticletitle{AMPS-Inf: Automatic Model Partitioning for
  Serverless Inference with Cost Efficiency}. In \bibinfo{booktitle}{\emph{50th
  International Conference on Parallel Processing}}. \bibinfo{pages}{1--12}.
\newblock


\bibitem[Jeon et~al\mbox{.}(2020)]%
        {socc_partition_model}
\bibfield{author}{\bibinfo{person}{Beomyeol Jeon}, \bibinfo{person}{Linda Cai},
  \bibinfo{person}{Pallavi Srivastava}, \bibinfo{person}{Jintao Jiang},
  \bibinfo{person}{Xiaolan Ke}, \bibinfo{person}{Yitao Meng},
  \bibinfo{person}{Cong Xie}, {and} \bibinfo{person}{Indranil Gupta}.}
  \bibinfo{year}{2020}\natexlab{}.
\newblock \showarticletitle{Baechi: fast device placement of machine learning
  graphs}. In \bibinfo{booktitle}{\emph{Proceedings of the 11th ACM Symposium
  on Cloud Computing}}. \bibinfo{pages}{416--430}.
\newblock


\bibitem[Jeon et~al\mbox{.}(2019)]%
        {philly}
\bibfield{author}{\bibinfo{person}{Myeongjae Jeon}, \bibinfo{person}{Shivaram
  Venkataraman}, \bibinfo{person}{Amar Phanishayee}, \bibinfo{person}{Junjie
  Qian}, \bibinfo{person}{Wencong Xiao}, {and} \bibinfo{person}{Fan Yang}.}
  \bibinfo{year}{2019}\natexlab{}.
\newblock \showarticletitle{Analysis of Large-Scale Multi-Tenant GPU Clusters
  for DNN Training Workloads}. In \bibinfo{booktitle}{\emph{2019 USENIX Annual
  Technical Conference (USENIX ATC 19)}}. \bibinfo{pages}{947--960}.
\newblock


\bibitem[Jia et~al\mbox{.}(2018)]%
        {tensor_parallelism2}
\bibfield{author}{\bibinfo{person}{Zhihao Jia}, \bibinfo{person}{Sina Lin},
  \bibinfo{person}{Charles~R. Qi}, {and} \bibinfo{person}{Alex Aiken}.}
  \bibinfo{year}{2018}\natexlab{}.
\newblock \showarticletitle{Exploring Hidden Dimensions in Parallelizing
  Convolutional Neural Networks}. In \bibinfo{booktitle}{\emph{Proceedings of
  the 35th International Conference on Machine Learning}},
  Vol.~\bibinfo{volume}{80}. \bibinfo{pages}{2279--2288}.
\newblock


\bibitem[Jia et~al\mbox{.}(2019)]%
        {hybrid_parallelism1}
\bibfield{author}{\bibinfo{person}{Zhihao Jia}, \bibinfo{person}{Matei
  Zaharia}, {and} \bibinfo{person}{Alex Aiken}.}
  \bibinfo{year}{2019}\natexlab{}.
\newblock \showarticletitle{Beyond Data and Model Parallelism for Deep Neural
  Networks.}
\newblock \bibinfo{journal}{\emph{Proceedings of Machine Learning and Systems}}
   \bibinfo{volume}{1} (\bibinfo{year}{2019}), \bibinfo{pages}{1--13}.
\newblock


\bibitem[Jiang et~al\mbox{.}(2021)]%
        {lambdaml}
\bibfield{author}{\bibinfo{person}{Jiawei Jiang}, \bibinfo{person}{Shaoduo
  Gan}, \bibinfo{person}{Yue Liu}, \bibinfo{person}{Fanlin Wang},
  \bibinfo{person}{Gustavo Alonso}, \bibinfo{person}{Ana Klimovic},
  \bibinfo{person}{Ankit Singla}, \bibinfo{person}{Wentao Wu}, {and}
  \bibinfo{person}{Ce Zhang}.} \bibinfo{year}{2021}\natexlab{}.
\newblock \showarticletitle{Towards demystifying serverless machine learning
  training}. In \bibinfo{booktitle}{\emph{Proceedings of the 2021 International
  Conference on Management of Data}}. \bibinfo{pages}{857--871}.
\newblock


\bibitem[Klimovic et~al\mbox{.}(2018a)]%
        {serverless_bandwidth1}
\bibfield{author}{\bibinfo{person}{Ana Klimovic}, \bibinfo{person}{Yawen Wang},
  \bibinfo{person}{Christos Kozyrakis}, \bibinfo{person}{Patrick Stuedi},
  \bibinfo{person}{Jonas Pfefferle}, {and} \bibinfo{person}{Animesh Trivedi}.}
  \bibinfo{year}{2018}\natexlab{a}.
\newblock \showarticletitle{Understanding ephemeral storage for serverless
  analytics}. In \bibinfo{booktitle}{\emph{2018 USENIX Annual Technical
  Conference (USENIX ATC 18)}}. \bibinfo{pages}{789--794}.
\newblock


\bibitem[Klimovic et~al\mbox{.}(2018b)]%
        {pocket}
\bibfield{author}{\bibinfo{person}{Ana Klimovic}, \bibinfo{person}{Yawen Wang},
  \bibinfo{person}{Patrick Stuedi}, \bibinfo{person}{Animesh Trivedi},
  \bibinfo{person}{Jonas Pfefferle}, {and} \bibinfo{person}{Christos
  Kozyrakis}.} \bibinfo{year}{2018}\natexlab{b}.
\newblock \showarticletitle{Pocket: Elastic ephemeral storage for serverless
  analytics}. In \bibinfo{booktitle}{\emph{13th USENIX Symposium on Operating
  Systems Design and Implementation (OSDI 18)}}. \bibinfo{pages}{427--444}.
\newblock


\bibitem[Li et~al\mbox{.}(2020a)]%
        {hyper-tuning-ns-1}
\bibfield{author}{\bibinfo{person}{Liam Li}, \bibinfo{person}{Kevin Jamieson},
  \bibinfo{person}{Afshin Rostamizadeh}, \bibinfo{person}{Ekaterina Gonina},
  \bibinfo{person}{Jonathan Ben-tzur}, \bibinfo{person}{Moritz Hardt},
  \bibinfo{person}{Benjamin Recht}, {and} \bibinfo{person}{Ameet Talwalkar}.}
  \bibinfo{year}{2020}\natexlab{a}.
\newblock \showarticletitle{A System for Massively Parallel Hyperparameter
  Tuning}. In \bibinfo{booktitle}{\emph{Proceedings of Machine Learning and
  Systems}}, Vol.~\bibinfo{volume}{2}. \bibinfo{pages}{230--246}.
\newblock


\bibitem[Li and Hoefler(2021)]%
        {chimera}
\bibfield{author}{\bibinfo{person}{Shigang Li} {and} \bibinfo{person}{Torsten
  Hoefler}.} \bibinfo{year}{2021}\natexlab{}.
\newblock \showarticletitle{Chimera: efficiently training large-scale neural
  networks with bidirectional pipelines}. In
  \bibinfo{booktitle}{\emph{Proceedings of the International Conference for
  High Performance Computing, Networking, Storage and Analysis}}.
  \bibinfo{pages}{1--14}.
\newblock


\bibitem[Li et~al\mbox{.}(2021)]%
        {li2021syncswitch_icdcs}
\bibfield{author}{\bibinfo{person}{Shijian Li}, \bibinfo{person}{Oren
  Mangoubi}, \bibinfo{person}{Lijie Xu}, {and} \bibinfo{person}{Tian Guo}.}
  \bibinfo{year}{2021}\natexlab{}.
\newblock \showarticletitle{Sync-Switch: Hybrid Parameter Synchronization for
  Distributed Deep Learning}. In \bibinfo{booktitle}{\emph{2021 IEEE 41th
  International Conference on Distributed Computing Systems (ICDCS)}}.
\newblock


\bibitem[Li et~al\mbox{.}(2020b)]%
        {li2019cmdare_icdcs}
\bibfield{author}{\bibinfo{person}{Shijian Li}, \bibinfo{person}{Robert~J.
  Walls}, {and} \bibinfo{person}{Tian Guo}.} \bibinfo{year}{2020}\natexlab{b}.
\newblock \showarticletitle{Characterizing and Modeling Distributed Training
  with Transient Cloud GPU Servers}. In \bibinfo{booktitle}{\emph{2020 IEEE
  40th International Conference on Distributed Computing Systems (ICDCS)}}.
\newblock


\bibitem[Li et~al\mbox{.}(2020c)]%
        {data_parallelism1}
\bibfield{author}{\bibinfo{person}{Shen Li}, \bibinfo{person}{Yanli Zhao},
  \bibinfo{person}{Rohan Varma}, \bibinfo{person}{Omkar Salpekar},
  \bibinfo{person}{Pieter Noordhuis}, \bibinfo{person}{Teng Li},
  \bibinfo{person}{Adam Paszke}, \bibinfo{person}{Jeff Smith},
  \bibinfo{person}{Brian Vaughan}, \bibinfo{person}{Pritam Damania}, {and}
  \bibinfo{person}{Soumith Chintala}.} \bibinfo{year}{2020}\natexlab{c}.
\newblock \showarticletitle{PyTorch Distributed: Experiences on Accelerating
  Data Parallel Training}.
\newblock \bibinfo{journal}{\emph{Proc. VLDB Endow.}} \bibinfo{volume}{13},
  \bibinfo{number}{12} (\bibinfo{year}{2020}), \bibinfo{pages}{3005–3018}.
\newblock


\bibitem[McDonald et~al\mbox{.}(2010)]%
        {asynchronous1}
\bibfield{author}{\bibinfo{person}{Ryan McDonald}, \bibinfo{person}{Keith
  Hall}, {and} \bibinfo{person}{Gideon Mann}.} \bibinfo{year}{2010}\natexlab{}.
\newblock \showarticletitle{Distributed training strategies for the structured
  perceptron}. In \bibinfo{booktitle}{\emph{Human language technologies: The
  2010 annual conference of the North American chapter of the association for
  computational linguistics}}. \bibinfo{pages}{456--464}.
\newblock


\bibitem[McGrath and Brenner(2017)]%
        {cold_start2}
\bibfield{author}{\bibinfo{person}{Garrett McGrath} {and}
  \bibinfo{person}{Paul~R Brenner}.} \bibinfo{year}{2017}\natexlab{}.
\newblock \showarticletitle{Serverless computing: Design, implementation, and
  performance}. In \bibinfo{booktitle}{\emph{2017 IEEE 37th International
  Conference on Distributed Computing Systems Workshops (ICDCSW)}}. IEEE,
  \bibinfo{pages}{405--410}.
\newblock


\bibitem[Microsoft(2022a)]%
        {azure_cloud}
\bibfield{author}{\bibinfo{person}{Microsoft}.}
  \bibinfo{year}{2022}\natexlab{a}.
\newblock \bibinfo{title}{Microsoft Azure Cloud Computing}.
\newblock
\newblock
\newblock
\shownote{\url{https://azure.microsoft.com/}}.


\bibitem[Microsoft(2022b)]%
        {azure_storage}
\bibfield{author}{\bibinfo{person}{Microsoft}.}
  \bibinfo{year}{2022}\natexlab{b}.
\newblock \bibinfo{title}{Microsoft Azure Storage}.
\newblock
\newblock
\newblock
\shownote{\url{https://azure.microsoft.com/services/storage}}.


\bibitem[Mirhoseini et~al\mbox{.}(2018)]%
        {iclr_partition_model}
\bibfield{author}{\bibinfo{person}{Azalia Mirhoseini}, \bibinfo{person}{Anna
  Goldie}, \bibinfo{person}{Hieu Pham}, \bibinfo{person}{Benoit Steiner},
  \bibinfo{person}{Quoc~V. Le}, {and} \bibinfo{person}{Jeff Dean}.}
  \bibinfo{year}{2018}\natexlab{}.
\newblock \showarticletitle{Hierarchical Planning for Device Placement}. In
  \bibinfo{booktitle}{\emph{ICLR}}.
\newblock


\bibitem[Mirhoseini et~al\mbox{.}(2017)]%
        {icml_partition_model}
\bibfield{author}{\bibinfo{person}{Azalia Mirhoseini}, \bibinfo{person}{Hieu
  Pham}, \bibinfo{person}{Quoc~V Le}, \bibinfo{person}{Benoit Steiner},
  \bibinfo{person}{Rasmus Larsen}, \bibinfo{person}{Yuefeng Zhou},
  \bibinfo{person}{Naveen Kumar}, \bibinfo{person}{Mohammad Norouzi},
  \bibinfo{person}{Samy Bengio}, {and} \bibinfo{person}{Jeff Dean}.}
  \bibinfo{year}{2017}\natexlab{}.
\newblock \showarticletitle{Device placement optimization with reinforcement
  learning}. In \bibinfo{booktitle}{\emph{International Conference on Machine
  Learning}}. PMLR, \bibinfo{pages}{2430--2439}.
\newblock


\bibitem[Narayanan et~al\mbox{.}(2019)]%
        {pipedream}
\bibfield{author}{\bibinfo{person}{Deepak Narayanan}, \bibinfo{person}{Aaron
  Harlap}, \bibinfo{person}{Amar Phanishayee}, \bibinfo{person}{Vivek
  Seshadri}, \bibinfo{person}{Nikhil~R Devanur}, \bibinfo{person}{Gregory~R
  Ganger}, \bibinfo{person}{Phillip~B Gibbons}, {and} \bibinfo{person}{Matei
  Zaharia}.} \bibinfo{year}{2019}\natexlab{}.
\newblock \showarticletitle{PipeDream: generalized pipeline parallelism for DNN
  training}. In \bibinfo{booktitle}{\emph{Proceedings of the 27th ACM Symposium
  on Operating Systems Principles}}. \bibinfo{pages}{1--15}.
\newblock


\bibitem[Narayanan et~al\mbox{.}(2021)]%
        {megatron2}
\bibfield{author}{\bibinfo{person}{Deepak Narayanan}, \bibinfo{person}{Mohammad
  Shoeybi}, \bibinfo{person}{Jared Casper}, \bibinfo{person}{Patrick
  LeGresley}, \bibinfo{person}{Mostofa Patwary}, \bibinfo{person}{Vijay
  Korthikanti}, \bibinfo{person}{Dmitri Vainbrand}, \bibinfo{person}{Prethvi
  Kashinkunti}, \bibinfo{person}{Julie Bernauer}, \bibinfo{person}{Bryan
  Catanzaro}, \bibinfo{person}{Amar Phanishayee}, {and} \bibinfo{person}{Matei
  Zaharia}.} \bibinfo{year}{2021}\natexlab{}.
\newblock \showarticletitle{Efficient Large-Scale Language Model Training on
  GPU Clusters Using Megatron-LM}. In \bibinfo{booktitle}{\emph{Proceedings of
  the International Conference for High Performance Computing, Networking,
  Storage and Analysis}}.
\newblock


\bibitem[Ngatchou et~al\mbox{.}(2005)]%
        {pareto}
\bibfield{author}{\bibinfo{person}{Patrick Ngatchou}, \bibinfo{person}{Anahita
  Zarei}, {and} \bibinfo{person}{A El-Sharkawi}.}
  \bibinfo{year}{2005}\natexlab{}.
\newblock \showarticletitle{Pareto multi objective optimization}. In
  \bibinfo{booktitle}{\emph{Proceedings of the 13th International Conference
  on, Intelligent Systems Application to Power Systems}}. IEEE,
  \bibinfo{pages}{84--91}.
\newblock


\bibitem[Patarasuk and Yuan(2007)]%
        {all_reduce3}
\bibfield{author}{\bibinfo{person}{Pitch Patarasuk} {and} \bibinfo{person}{Xin
  Yuan}.} \bibinfo{year}{2007}\natexlab{}.
\newblock \showarticletitle{Bandwidth efficient all-reduce operation on tree
  topologies}. In \bibinfo{booktitle}{\emph{2007 IEEE International Parallel
  and Distributed Processing Symposium}}. IEEE, \bibinfo{pages}{1--8}.
\newblock


\bibitem[Patarasuk and Yuan(2009)]%
        {all_reduce2}
\bibfield{author}{\bibinfo{person}{Pitch Patarasuk} {and} \bibinfo{person}{Xin
  Yuan}.} \bibinfo{year}{2009}\natexlab{}.
\newblock \showarticletitle{Bandwidth optimal all-reduce algorithms for
  clusters of workstations}.
\newblock \bibinfo{journal}{\emph{J. Parallel and Distrib. Comput.}}
  \bibinfo{volume}{69}, \bibinfo{number}{2} (\bibinfo{year}{2009}),
  \bibinfo{pages}{117--124}.
\newblock


\bibitem[Rausch et~al\mbox{.}(2019)]%
        {serverless_application3}
\bibfield{author}{\bibinfo{person}{Thomas Rausch}, \bibinfo{person}{Waldemar
  Hummer}, \bibinfo{person}{Vinod Muthusamy}, \bibinfo{person}{Alexander
  Rashed}, {and} \bibinfo{person}{Schahram Dustdar}.}
  \bibinfo{year}{2019}\natexlab{}.
\newblock \showarticletitle{Towards a serverless platform for edge {AI}}. In
  \bibinfo{booktitle}{\emph{2nd USENIX Workshop on Hot Topics in Edge Computing
  (HotEdge 19)}}.
\newblock


\bibitem[Research(2017)]%
        {ringallreduce}
\bibfield{author}{\bibinfo{person}{Baidu Research}.}
  \bibinfo{year}{2017}\natexlab{}.
\newblock \bibinfo{title}{baidu-allreduce}.
\newblock
\newblock
\newblock
\shownote{\url{https://github.com/baidu-research/baidu-allreduce}}.


\bibitem[Romero et~al\mbox{.}(2021a)]%
        {Romero2021-ly}
\bibfield{author}{\bibinfo{person}{Francisco Romero},
  \bibinfo{person}{Gohar~Irfan Chaudhry}, \bibinfo{person}{{\'I}{\~n}igo
  Goiri}, \bibinfo{person}{Pragna Gopa}, \bibinfo{person}{Paul Batum},
  \bibinfo{person}{Neeraja~J Yadwadkar}, \bibinfo{person}{Rodrigo Fonseca},
  \bibinfo{person}{Christos Kozyrakis}, {and} \bibinfo{person}{Ricardo
  Bianchini}.} \bibinfo{year}{2021}\natexlab{a}.
\newblock \showarticletitle{{Faa\$T: A Transparent Auto-Scaling Cache for
  Serverless Applications}}.
\newblock  (\bibinfo{year}{2021}).
\newblock


\bibitem[Romero et~al\mbox{.}(2021b)]%
        {llama}
\bibfield{author}{\bibinfo{person}{Francisco Romero}, \bibinfo{person}{Mark
  Zhao}, \bibinfo{person}{Neeraja~J. Yadwadkar}, {and}
  \bibinfo{person}{Christos Kozyrakis}.} \bibinfo{year}{2021}\natexlab{b}.
\newblock \showarticletitle{Llama: A Heterogeneous \& Serverless Framework for
  Auto-Tuning Video Analytics Pipelines}. In
  \bibinfo{booktitle}{\emph{Proceedings of the ACM Symposium on Cloud
  Computing}} (Seattle, WA, USA) \emph{(\bibinfo{series}{SoCC'21})}.
\newblock


\bibitem[Shallue et~al\mbox{.}(2019)]%
        {data_parallelism2}
\bibfield{author}{\bibinfo{person}{Christopher~J. Shallue},
  \bibinfo{person}{Jaehoon Lee}, \bibinfo{person}{Joseph Antognini},
  \bibinfo{person}{Jascha Sohl-Dickstein}, \bibinfo{person}{Roy Frostig}, {and}
  \bibinfo{person}{George~E. Dahl}.} \bibinfo{year}{2019}\natexlab{}.
\newblock \showarticletitle{Measuring the Effects of Data Parallelism on Neural
  Network Training}.
\newblock \bibinfo{journal}{\emph{Journal of Machine Learning Research}}
  \bibinfo{volume}{20}, \bibinfo{number}{112} (\bibinfo{year}{2019}),
  \bibinfo{pages}{1--49}.
\newblock


\bibitem[Shazeer et~al\mbox{.}(2018)]%
        {tensor_parallelism1}
\bibfield{author}{\bibinfo{person}{Noam Shazeer}, \bibinfo{person}{Youlong
  Cheng}, \bibinfo{person}{Niki Parmar}, \bibinfo{person}{Dustin Tran},
  \bibinfo{person}{Ashish Vaswani}, \bibinfo{person}{Penporn Koanantakool},
  \bibinfo{person}{Peter Hawkins}, {et~al\mbox{.}}}
  \bibinfo{year}{2018}\natexlab{}.
\newblock \showarticletitle{Mesh-tensorflow: Deep learning for supercomputers}.
\newblock \bibinfo{journal}{\emph{Advances in neural information processing
  systems}}  \bibinfo{volume}{31} (\bibinfo{year}{2018}).
\newblock


\bibitem[Shoeybi et~al\mbox{.}(2019)]%
        {megatron}
\bibfield{author}{\bibinfo{person}{Mohammad Shoeybi}, \bibinfo{person}{Mostofa
  Patwary}, \bibinfo{person}{Raul Puri}, \bibinfo{person}{Patrick LeGresley},
  \bibinfo{person}{Jared Casper}, {and} \bibinfo{person}{Bryan Catanzaro}.}
  \bibinfo{year}{2019}\natexlab{}.
\newblock \showarticletitle{Megatron-lm: Training multi-billion parameter
  language models using model parallelism}.
\newblock \bibinfo{journal}{\emph{arXiv preprint arXiv:1909.08053}}
  (\bibinfo{year}{2019}).
\newblock


\bibitem[Sohoni et~al\mbox{.}(2019)]%
        {ga1}
\bibfield{author}{\bibinfo{person}{Nimit~Sharad Sohoni},
  \bibinfo{person}{Christopher~Richard Aberger}, \bibinfo{person}{Megan
  Leszczynski}, \bibinfo{person}{Jian Zhang}, {and}
  \bibinfo{person}{Christopher R{\'e}}.} \bibinfo{year}{2019}\natexlab{}.
\newblock \showarticletitle{Low-memory neural network training: A technical
  report}.
\newblock \bibinfo{journal}{\emph{arXiv preprint arXiv:1904.10631}}
  (\bibinfo{year}{2019}).
\newblock


\bibitem[Song et~al\mbox{.}(2020)]%
        {ga2}
\bibfield{author}{\bibinfo{person}{Liuyihan Song}, \bibinfo{person}{Pan Pan},
  \bibinfo{person}{Kang Zhao}, \bibinfo{person}{Hao Yang},
  \bibinfo{person}{Yiming Chen}, \bibinfo{person}{Yingya Zhang},
  \bibinfo{person}{Yinghui Xu}, {and} \bibinfo{person}{Rong Jin}.}
  \bibinfo{year}{2020}\natexlab{}.
\newblock \showarticletitle{Large-scale training system for 100-million
  classification at alibaba}. In \bibinfo{booktitle}{\emph{Proceedings of the
  26th ACM SIGKDD International Conference on Knowledge Discovery \& Data
  Mining}}. \bibinfo{pages}{2909--2930}.
\newblock


\bibitem[Tarnawski et~al\mbox{.}(2020)]%
        {nips_partition}
\bibfield{author}{\bibinfo{person}{Jakub~M Tarnawski}, \bibinfo{person}{Amar
  Phanishayee}, \bibinfo{person}{Nikhil Devanur}, \bibinfo{person}{Divya
  Mahajan}, {and} \bibinfo{person}{Fanny Nina~Paravecino}.}
  \bibinfo{year}{2020}\natexlab{}.
\newblock \showarticletitle{Efficient algorithms for device placement of dnn
  graph operators}.
\newblock \bibinfo{journal}{\emph{Advances in Neural Information Processing
  Systems}}  \bibinfo{volume}{33} (\bibinfo{year}{2020}),
  \bibinfo{pages}{15451--15463}.
\newblock


\bibitem[Thakur et~al\mbox{.}(2005)]%
        {all_reduce1}
\bibfield{author}{\bibinfo{person}{Rajeev Thakur}, \bibinfo{person}{Rolf
  Rabenseifner}, {and} \bibinfo{person}{William Gropp}.}
  \bibinfo{year}{2005}\natexlab{}.
\newblock \showarticletitle{Optimization of collective communication operations
  in MPICH}.
\newblock \bibinfo{journal}{\emph{The International Journal of High Performance
  Computing Applications}} \bibinfo{volume}{19}, \bibinfo{number}{1}
  (\bibinfo{year}{2005}), \bibinfo{pages}{49--66}.
\newblock


\bibitem[Thorpe et~al\mbox{.}(2021)]%
        {dorylus}
\bibfield{author}{\bibinfo{person}{John Thorpe}, \bibinfo{person}{Yifan Qiao},
  \bibinfo{person}{Jonathan Eyolfson}, \bibinfo{person}{Shen Teng},
  \bibinfo{person}{Guanzhou Hu}, \bibinfo{person}{Zhihao Jia},
  \bibinfo{person}{Jinliang Wei}, \bibinfo{person}{Keval Vora},
  \bibinfo{person}{Ravi Netravali}, \bibinfo{person}{Miryung Kim},
  {et~al\mbox{.}}} \bibinfo{year}{2021}\natexlab{}.
\newblock \showarticletitle{Dorylus: Affordable, Scalable, and Accurate {GNN}
  Training with Distributed {CPU} Servers and Serverless Threads}. In
  \bibinfo{booktitle}{\emph{15th USENIX Symposium on Operating Systems Design
  and Implementation (OSDI 21)}}. \bibinfo{pages}{495--514}.
\newblock


\bibitem[Vahidinia et~al\mbox{.}(2020)]%
        {cold_start1}
\bibfield{author}{\bibinfo{person}{Parichehr Vahidinia}, \bibinfo{person}{Bahar
  Farahani}, {and} \bibinfo{person}{Fereidoon~Shams Aliee}.}
  \bibinfo{year}{2020}\natexlab{}.
\newblock \showarticletitle{Cold start in serverless computing: Current trends
  and mitigation strategies}. In \bibinfo{booktitle}{\emph{2020 International
  Conference on Omni-layer Intelligent Systems (COINS)}}. IEEE,
  \bibinfo{pages}{1--7}.
\newblock


\bibitem[Wang et~al\mbox{.}(2020b)]%
        {Wang2020-to}
\bibfield{author}{\bibinfo{person}{Ao Wang}, \bibinfo{person}{Jingyuan Zhang},
  \bibinfo{person}{Xiaolong Ma}, \bibinfo{person}{Ali Anwar},
  \bibinfo{person}{Lukas Rupprecht}, \bibinfo{person}{Dimitrios Skourtis},
  \bibinfo{person}{Vasily Tarasov}, \bibinfo{person}{Feng Yan}, {and}
  \bibinfo{person}{Yue Cheng}.} \bibinfo{year}{2020}\natexlab{b}.
\newblock \showarticletitle{{InfiniCache: Exploiting Ephemeral Serverless
  Functions to Build a Cost-Effective Memory Cache}}. In
  \bibinfo{booktitle}{\emph{{18th {USENIX} Conference on File and Storage
  Technologies ({FAST} 20)}}}. \bibinfo{publisher}{usenix.org},
  \bibinfo{pages}{267--281}.
\newblock


\bibitem[Wang et~al\mbox{.}(2019)]%
        {siren}
\bibfield{author}{\bibinfo{person}{Hao Wang}, \bibinfo{person}{Di Niu}, {and}
  \bibinfo{person}{Baochun Li}.} \bibinfo{year}{2019}\natexlab{}.
\newblock \showarticletitle{Distributed machine learning with a serverless
  architecture}. In \bibinfo{booktitle}{\emph{IEEE INFOCOM 2019-IEEE Conference
  on Computer Communications}}. IEEE, \bibinfo{pages}{1288--1296}.
\newblock


\bibitem[Wang et~al\mbox{.}(2020a)]%
        {linnanlamcts}
\bibfield{author}{\bibinfo{person}{Linnan Wang}, \bibinfo{person}{Rodrigo
  Fonseca}, {and} \bibinfo{person}{Yuandong Tian}.}
  \bibinfo{year}{2020}\natexlab{a}.
\newblock \showarticletitle{Learning Search Space Partition for Black-box
  Optimization using Monte Carlo Tree Search}. In
  \bibinfo{booktitle}{\emph{Advances in Neural Information Processing Systems
  (NeurIPS), 2020}} (online).
\newblock


\bibitem[Wang et~al\mbox{.}(2018)]%
        {serverless_bandwidth2}
\bibfield{author}{\bibinfo{person}{Liang Wang}, \bibinfo{person}{Mengyuan Li},
  \bibinfo{person}{Yinqian Zhang}, \bibinfo{person}{Thomas Ristenpart}, {and}
  \bibinfo{person}{Michael Swift}.} \bibinfo{year}{2018}\natexlab{}.
\newblock \showarticletitle{Peeking behind the curtains of serverless
  platforms}. In \bibinfo{booktitle}{\emph{2018 USENIX Annual Technical
  Conference (USENIX ATC 18)}}. \bibinfo{pages}{133--146}.
\newblock


\bibitem[Wawrzoniak and Bruno(2021)]%
        {nat3}
\bibfield{author}{\bibinfo{person}{Ingo Wawrzoniak, Mike} {and}
  \bibinfo{person}{Fraga Barcelos~Paulus Bruno}.}
  \bibinfo{year}{2021}\natexlab{}.
\newblock \showarticletitle{Boxer: Data Analytics on Network-enabled Serverless
  Platforms}. In \bibinfo{booktitle}{\emph{11th Annual Conference on Innovative
  Data Systems Research}}.
\newblock


\bibitem[Xu et~al\mbox{.}(2021)]%
        {lambdadnn}
\bibfield{author}{\bibinfo{person}{Fei Xu}, \bibinfo{person}{Yiling Qin},
  \bibinfo{person}{Li Chen}, \bibinfo{person}{Zhi Zhou}, {and}
  \bibinfo{person}{Fangming Liu}.} \bibinfo{year}{2021}\natexlab{}.
\newblock \showarticletitle{$\lambda$DNN: Achieving Predictable Distributed DNN
  Training With Serverless Architectures}.
\newblock \bibinfo{journal}{\emph{IEEE Trans. Comput.}} \bibinfo{volume}{71},
  \bibinfo{number}{2} (\bibinfo{year}{2021}), \bibinfo{pages}{450--463}.
\newblock


\bibitem[Yu et~al\mbox{.}(2021)]%
        {gillis}
\bibfield{author}{\bibinfo{person}{Minchen Yu}, \bibinfo{person}{Zhifeng
  Jiang}, {et~al\mbox{.}}} \bibinfo{year}{2021}\natexlab{}.
\newblock \showarticletitle{Gillis: Serving Large Neural Networks in Serverless
  Functions with Automatic Model Partitioning}. In
  \bibinfo{booktitle}{\emph{2021 IEEE 41st International Conference on
  Distributed Computing Systems (ICDCS)}}. IEEE, \bibinfo{pages}{138--148}.
\newblock


\bibitem[Zhang et~al\mbox{.}(2019)]%
        {shredder}
\bibfield{author}{\bibinfo{person}{Tian Zhang}, \bibinfo{person}{Dong Xie},
  \bibinfo{person}{Feifei Li}, {and} \bibinfo{person}{Ryan Stutsman}.}
  \bibinfo{year}{2019}\natexlab{}.
\newblock \showarticletitle{Narrowing the gap between serverless and its state
  with storage functions}. In \bibinfo{booktitle}{\emph{Proceedings of the ACM
  Symposium on Cloud Computing}}. \bibinfo{pages}{1--12}.
\newblock


\bibitem[Zhang et~al\mbox{.}(2016)]%
        {asynchronous2}
\bibfield{author}{\bibinfo{person}{Wei Zhang}, \bibinfo{person}{Suyog Gupta},
  \bibinfo{person}{Xiangru Lian}, {and} \bibinfo{person}{Ji Liu}.}
  \bibinfo{year}{2016}\natexlab{}.
\newblock \showarticletitle{Staleness-Aware Async-SGD for Distributed Deep
  Learning}. In \bibinfo{booktitle}{\emph{Proceedings of the Twenty-Fifth
  International Joint Conference on Artificial Intelligence}}
  \emph{(\bibinfo{series}{IJCAI'16})}. \bibinfo{pages}{2350–2356}.
\newblock


\bibitem[Zhang et~al\mbox{.}(2020)]%
        {distributed_network}
\bibfield{author}{\bibinfo{person}{Zhen Zhang}, \bibinfo{person}{Chaokun
  Chang}, \bibinfo{person}{Haibin Lin}, \bibinfo{person}{Yida Wang},
  \bibinfo{person}{Raman Arora}, {and} \bibinfo{person}{Xin Jin}.}
  \bibinfo{year}{2020}\natexlab{}.
\newblock \showarticletitle{Is network the bottleneck of distributed
  training?}. In \bibinfo{booktitle}{\emph{Proceedings of the Workshop on
  Network Meets AI \& ML}}. \bibinfo{pages}{8--13}.
\newblock


\bibitem[Zhao et~al\mbox{.}(2022)]%
        {Zhao2021MultiobjectiveOB}
\bibfield{author}{\bibinfo{person}{Yiyang Zhao}, \bibinfo{person}{Linnan Wang},
  \bibinfo{person}{Kevin Yang}, \bibinfo{person}{Tianjun Zhang},
  \bibinfo{person}{Tian Guo}, {and} \bibinfo{person}{Yuandong Tian}.}
  \bibinfo{year}{2022}\natexlab{}.
\newblock \showarticletitle{Multi-objective Optimization by Learning Space
  Partitions}.
\newblock \bibinfo{journal}{\emph{10th International Conference on Learning
  Representations, {ICLR}}} (\bibinfo{year}{2022}).
\newblock


\end{thebibliography}
